\documentclass[useAMS,usenatbib, 
]{mnras}
\usepackage{natbib}
\usepackage{graphicx}
\usepackage{epsfig}
\usepackage{amssymb}
\usepackage{amsmath}
\usepackage{natbib}
\usepackage{times}

\usepackage{comment}

\usepackage{bm}

\usepackage[english]{babel}

\title[Explosive instability of gas-dust mixture]{Explosive instability of dust settling in a protoplanetary disc}
\author[V.V. Zhuravlev]{V. V. Zhuravlev$^{1}$\thanks{E-mail:
zhuravlev@sai.msu.ru} \\
$^{1}$Sternberg Astronomical Institute, Lomonosov Moscow State University, Universitetskij pr., 13, Moscow 119234, Russia}

\begin{document}

\date{
}

\pagerange{\pageref{firstpage}--\pageref{lastpage}} \pubyear{2017}

\maketitle

\label{firstpage}

\defcitealias{zhuravlev-2019}{Z19}

\begin{abstract}

It is shown that gas-dust perturbations in a disc with dust settling to the disc midplane exhibit the non-linear 
three-wave resonant interactions between streaming dust wave (SDW) and two inertial waves (IW). 
In the particular case considered in this paper, SDW at the wavenumber 
$k^\bullet = 2\kappa / (g_z t_s)$, where $\kappa$, $g_z$ and $t_s$ are, respectively, epicyclic frequency, vertical gravitational acceleration and particle's stopping time, interacts with two IW at the lower wavenumbers $k^\prime$ and $k^{\prime\prime}$ such that $k^\prime < k_{\rm DSI} < k^{\prime\prime} < k^\bullet$, where 
$k_{\rm DSI} = \kappa / (g_z t_s)$ is the wavenumber of the linear resonance between SDW and IW associated with 
the previously discovered linear dust settling instability. 
The problem is solved analytically in the limit of the small dust fraction. 
As soon as the dynamical dust back reaction on gas is taken into account, ${\bf k}^\bullet$, 
${\bf k}^\prime$ and ${\bf k}^{\prime\prime}$ become slightly non-collinear and the emerging 
interaction of waves leads to simultaneous explosive growth of their amplitudes.
This growth is explained by the conservative exchange with energy between the waves. The amplitudes of all three
waves grow because the negative energy SDW transfers its energy to the positive energy IW.
The product of the dimensionless amplitude of initially dominant wave and the time of explosion 
can be less than Keplerian time in a disc. 
It is shown that, generally, the three-wave resonance of an explosive type exists in a wide range of wavenumbers $0 < k^\bullet \leq 2\kappa / (g_z t_s)$.
An explosive instability of gas-dust mixture may facilitate the dust clumping and the subsequent formation of 
planetesimals in young protoplanetary discs.

\end{abstract}


\section{Introduction}

The streaming instability of gas-dust mixture associated with the dust radial drift in the midplane of 
protoplanetary disc has been discovered by \citet{youdin-goodman-2005} and further extensively investigated 
using the numerical simulations \citep{johansen-2007-apj-1, johansen-2007-apj-2, johansen-2009, carrera-2015,
johansen-2017}.
The ability of the streaming instability to concentrate solids into high densities in the non-linear regime 
is accepted to be necessary for the formation of planetesimals. 
However, it has thresholds for small solids and low metallicities, which quantitatively depend on the details 
of numerical setup \citep{yang-2014, johansen-2017, li-youdin-2018, li-youdin-2021}.

The amount of dust in the disc midplane can be probably increased through the dust settling instability (DSI)
discovered by \citet{squire_2018}. DSI is caused by vertical rather than radial drift of the grains.
So far, the only attempt to study the dynamics of finite amplitude perturbations presumably associated with 
non-linear stage of DSI has been made by \citet{krapp-youdin-2020}. 
The numerical simulations performed with the multi-fluid code have shown either a weak or slow particles clumping
with a caveat about the convergence of the maximum dust density.


This work is another effort to study the non-linear dynamics of gas-dust perturbations with the account of the 
dust back reaction on gas, which takes place in a disc with the dust settling to its midplane. 
Unlike most of the previous work on the non-linear gas-dust dynamics 
in protoplanetary discs, it employs essentially the analytical approach. 
Of course, the latter becomes possible due to several simplifications of the considered problem. 
First, perturbations are considered in a patch of disc much smaller than the disc scaleheight. 
It is assumed that perturbations are axisymmetric, while the dust behaves like the second pressureless fluid. 
The dust is coupled with the gas through the drag force parametrised by the stopping time of the particles \citep{squire_2018}.
Further, the model is restricted by the case of the small dust fraction. 
Finally, the major point is that two-fluid dynamics is considered within the weakly
non-linear theory with only the quadratic interactions between the modes retained \citep{craik_book}.
As far as the amplitudes of perturbations are small, the quadratic interactions are most efficient for modes 
satisfying the three-wave resonant conditions \citep{kadomtsev-1971}.


The three-wave resonant interactions are responsible for redistribution of energy over the different scales
in various geophysical and astrophysical flows. They serve as main process coupling waves in a weak turbulence regime. 
The general theory of such a dynamics was introduced in physics of fluids by \citet{phillips-1960} who 
was the first to search for three-wave resonance among surface gravity waves. This work was followed by
many studies of three-wave resonance in various flows. 
For example, the three-wave resonance of capillary-gravity waves was investigated by \citet{mcgoldrick-1965}, 
see also its recent experimental verification by \citet{haudin-2016}.
The existence of resonant triads among three internal gravity waves, or alternatively,
among two surface gravity waves and one internal gravity wave in stratified medium 
has been shown by \citet{thorpe-1966}.
For the corresponding experimental study see, e.g., \citet{joubaud-2012} who performed the first measurement of 
the parametric subharmonic instability growth rate in a tank filled with stratified salt water.


Like internal waves, inertial waves (IW hereafter) propagating in a rotating fluid obey a similar 
anisotropic dispersion relation, which makes the frequency depend on the direction of propagation 
of wave rather than on its wavelength. Accordingly, it has been verified experimentally by \citet{bordes-2012}
that subharmonic secondary waves excited due to three-wave resonance among IW propagate closer to the
plane of rotation as compared with propagation of the primary wave. 
This intrinsic feature of resonance between plane IW gives rise to the anisotropic 
turbulent transfer of energy 
mainly in the direction perpendicular to the rotation axis and generation of columnar vortices in weakly
turbulent rotating flows, see e.g. \citet{smith-waleffe-1999} and \citet{galtier-2003}.


There has been also much work on the three-wave resonances in stellar interiors.
\citet{vandakurov-1965} found the possibility of converting the radial pulsation of the star into 
two non-radial modes with a sum of frequencies close to the frequency of the radial pulsation. 
Later on, \citet{dziembowski-1982} constructed the general theory of resonant interactions of stellar
perturbations. He argued that three-wave resonance may be a mechanism limiting the amplitudes of modes.
The variant of the theory of resonant interactions of stellar perturbations applied to distribution of modes 
having random phases has been developed by \citet{goldreich-1989}, for the most recent example of work on
such a problem for the red giants see \citet{weinberg-2021}.

Three-wave resonance has a striking manifestation in non-equilibrium media. 
Since the work on the interaction of waves in plasma penetrated by a beam of charged particles or 
electrostatic waves propagating in magnetised plasma, see \cite{dikasov-1965} and further \citet{CRS-1969}, 
\citet{fukai-1970}, it has been known that the resonant interaction 
of positive and negative energy waves may lead to amplitudes of all waves growing up to infinity at finite time. 
Such kind of solutions has been referred to as an explosive instability.
Afterwards, it has been detected in the laboratory plasma, see \citet{nakamura-1977} and \citet{sugaya-1978}.
However, explosive instability of beam-plasma system proved difficult to investigate in laboratory. 
An illustrative example of explosive instability can be found in \citet{fukai-1979}, who studied its evolution 
employing one-dimensional cold fluid model of an electron beam as well as the corresponding particle-in-cell simulations, 
see e.g. Figure 3 of their paper. Perfect agreement of the fluid code solution with the growth of perturbations 
predicted by the standard mode coupling equations for the corresponding resonant triad implied that the higher order 
wave couplings did not saturate the instability. Saturation was revealed in particle-in-cell simulations, which
demonstrated that the high enough perturbations cause mixing of electrons belonging to the beam and the surrounding 
plasma. This causes the heating of medium and the subsequent frequency mismatch in the resonant triad.

For the non-linear dynamical system of quite a general form, see 
\citet{dougherty-1970}, \citet{davidson-1972} and \citet{rabinovich-1973}, 
it was shown that for explosive instability to occur two conditions must be satisfied. 
First, one of the resonant waves must have the energy sign different from the energy signs of the other two waves.
Second, this wave must take the highest frequency in the absolute value comparing with the frequencies 
of the other two waves. \citet{cairns-1979} pointed out that these conditions are met for three-wave
resonance of waves propagating in a three-layer flow with step-wise profiles of density and 
velocity. Shortly after that, \citet{craik-adam-1979} confirmed this prediction of explosive instability by the direct 
calculation of the corresponding interaction coefficients. 

Explosive instability has been suggested to be responsible for some transient effects in terrestrial and space environment. 
For example, the waves of vorticity observed in alongshore oceanic currents can be generated by an explosive interactions 
within the corresponding resonant triads below the low-frequency threshold predicted by the linear stability theory but not seen in 
observational data, see \citet{shrira-1997}.
Explosive instability of kink waves existing in magnetic flux tubes, see \citet{ryutova-1988}, 
can manifest itself in quiescent prominences of solar atmosphere. High-resolution space observations of Sun reveal the 
growing ripples at the prominence/corona interface, which end up with rapid formation of mushroom-like disturbances. 
The sudden formation of such structures can be naturally explained by an explosive growth of negative energy kink waves 
when magnetised flow is stable with respect to the linear Kelvin-Helmholtz instability, see \citet{ryutova-2000} and \citet{ryutova-2010}.

This work considers the non-linear stability of dust settling through the gas being in vertical hydrostatic 
equilibrium at some height above the protoplanetary disc midplane. Previously, \citet{zhuravlev-2019},
hereafter \citetalias{zhuravlev-2019}, revealed 
that the linear perturbations of the dust density advected by the settling dust can be considered as the negative 
energy wave. This wave was referred to as the streaming dust wave (SDW hereafter). 
Its linear resonance with the positive energy IW propagating
in the gas gives rise to DSI. Whether an explosive three-wave resonant interaction among SDW and IW is possible 
on the same background is an issue addressed in this study. 
At the same time, it is important to note that such an instability must be absent in the case of radial drift of the dust 
settled to the disc midplane because then there is no negative energy SDW, see \citetalias{zhuravlev-2019}.

In Section \ref{sec_eqs} the dynamical equations for perturbations of gas-dust mixture 
are derived retaining the terms that are quadratic over the dimensionless amplitudes of perturbations. 
The reasonable assumptions of the small dust fraction and the small stopping time of the particles make the linear
problem analytically tractable with the only additional restriction that the solution is sought sufficiently 
far from the linear resonance between SDW and IW. This is exposed in Section \ref{sec_linear}.
The particular case of three-wave resonance between one SDW and two IW satisfying the general conditions 
of explosive instability is proposed in Section \ref{sec_res}.
In Section \ref{sec_coupling} an explosive interaction between these waves is derived using the linear solution at the 
corresponding frequencies obtained previously. A conservative type of wave interactions is checked in Section 
\ref{sec_energy}. At last, Section \ref{sec_t_e} is assigned for various estimates of the time of explosion. 
It is shown that explosive instability of gas-dust mixture can emerge at the physically reasonable time, which 
is much shorter than the timescale of the dust settling.

\section{Non-linear equations for dynamics of gas-dust mixture in a disc}
\label{sec_eqs}

Starting point of the present analysis is the set of two-fluid equations 
describing the local axisymmetric dynamics of a partially coupled gas-dust 
mixture in a protoplanetary disc with the dust back reaction on gas taken into account, 
see \citetalias{zhuravlev-2019}:

\begin{equation}
\label{eq_U}
\begin{aligned}
\partial_t {\bf U} - 2\Omega_0 U_y {\bf e}_x + (2-q) \Omega_0 U_x {\bf e}_y + ({\bf U}\nabla) {\bf U} = \\ 
\frac{\nabla p_0}{\rho_g} - \frac{\nabla(p+p_0)}{\rho},
\end{aligned}
\end{equation}

\begin{equation}
\label{eq_TVA}
\frac{\nabla (p+p_0)}{\rho} = \frac{{\bf V}}{t_s}.
\end{equation}

\begin{equation}
\label{eq_rho_g}
\nabla \cdot \left ( {\bf U} - \frac{\rho_p}{\rho}{\bf V} \right ) = 0,
\end{equation}

\begin{equation}
\label{eq_rho_tot}
\partial_t \rho_p + \nabla ( \rho {\bf U} ) = 0.
\end{equation}
The notations for important variables are summarised in the Appendix \ref{app_symb}. 
Equations (\ref{eq_U}-\ref{eq_rho_tot}) are written in terms of the centre-of-mass velocity, 
$$
{\bf U} \equiv \frac{\rho_g {\bf U}_g + \rho_p {\bf U}_p}{\rho},
$$
where ${\bf U}_{g,p}$ and $\rho_{g,p}$ are velocity and density of gas and dust, respectively, while $\rho \equiv \rho_g+\rho_p$ is the total density of mixture. It is assumed that gas and dust velocities are measured with respect to 
the reference shear velocity ${\bf U}_0 = -q \Omega_0 x {\bf e}_y$ as defined in \citetalias{zhuravlev-2019}, 
where $\Omega_0$ is angular velocity of the reference frame comoving with the certain patch of disc. 
The shear rate, $q$, takes approximately the Keplerian value, $q=3/2$, in protoplanetary discs.
The gas pressure, $p$, is measured with respect to the reference pressure $p_0$, which defines ${\bf U}_0$ through the following equations
\begin{equation}
\label{rad_0}
\frac{1}{\rho_g}\frac{\partial p_0}{\partial x} = -\frac{\partial \Phi}{\partial x} + \Omega_0^2 (r_0+x) + 2\Omega_0 U_0,
\end{equation}
\begin{equation}
\label{vert_0}
\frac{1}{\rho_g}\frac{\partial p_0}{\partial z} = -\frac{\partial \Phi}{\partial z},
\end{equation}
where $\Phi$ is the gravitational potential of the host star.

Equation (\ref{eq_TVA}) represents the marginal case of particles tightly coupled to the gas so that there is 
no time-lag of their dynamical response to the change of the gas acceleration. This limit is usually referred to as the terminal velocity approximation, see \citet{youdin-goodman-2005}.
It is known that for the terminal velocity approximation 
to be valid the stopping time of the particles, $t_s$, must be so small that 
\begin{equation}
\label{TVA_1}
\tau_* \equiv t_s \max\{t_{ev}^{-1}, \Omega_0 \} \ll 1,
\end{equation}
and
\begin{equation}
\label{TVA_2}
\frac{g_z t_s^2}{l_{ev}} \ll 1,
\end{equation}
where $t_{ev}$ and $l_{ev}$ are, respectively, the characteristic time- and length-scales of gas-dust mixture
dynamics, see the discussion of the general equations in \citetalias{zhuravlev-2019}.
An important parameter to be used below is the dimensionless stopping time akin to $\tau_*$
\begin{equation}
\label{tau_def}
\tau \equiv t_s \Omega_0.
\end{equation}

\subsection{Stationary solution}

The dust settling is described by the following solution of equations (\ref{eq_U}-\ref{eq_rho_tot}), see \citetalias{zhuravlev-2019}:

\begin{equation}
\label{bg_U}
{\bf U} = 0,
\end{equation}
\begin{equation}
\label{bg_p}
\frac{\nabla (p+p_0)}{\rho} = {\bf g},
\end{equation}

\begin{equation}
\label{bg_V}
{\bf V} = t_s {\bf g},
\end{equation}
where ${\bf g} = - g_z {\bf e}_z$ is the vertical gravitational acceleration.
Hence, the radial drift of the particles is neglected in this study.

Finally, equation (\ref{eq_rho_g}) combined with equation (\ref{bg_V}) implies that
\begin{equation}
\label{bg_sigma}
\rho_p=const
\end{equation}
on the local scale considered here.

\subsection{Weakly non-linear equations for axisymmetric gas-dust perturbations}

The Eulerian perturbation of the centre-of-mass velocity, ${\bf u}$, 
the Eulerian perturbation of enthalpy of gas-dust mixture, $W \equiv p^\prime / \rho$,  
where $p^\prime$ is the Eulerian perturbation of gas pressure, and the relative perturbation of the dust density, $\delta \equiv \rho_p^\prime / \rho_p$, where $\rho_p^\prime$ is the Eulerian perturbation of the dust density, are imposed on the background given by equations (\ref{bg_U}-\ref{bg_sigma}). These perturbations obey the following weakly non-linear equations
\begin{equation}
\label{eq_1}
\partial_t {\bf u} - 2 u_y {\bf e}_x + \frac{\tilde\kappa^2}{2} u_x {\bf e}_y + 
({\bf u} \cdot \nabla) {\bf u} = -\nabla W - \frac{f}{\tau} \delta {\bf e}_z + f \delta \nabla W,
\end{equation}

\begin{equation}
\label{eq_2}
\partial_t \delta = - 2 \tau \partial_x u_y + (1-f) \partial_z \delta + 
\tau \sum_{i,k=1}^2 \partial_k u_i \partial_i u_k - 
\nabla \cdot (\delta {\bf u}) + f \partial_z \delta^2.
\end{equation} 
The sum in the right-hand side (RHS) of equation (\ref{eq_2}) is done over $x$- and $z$-projections of the velocity 
perturbation. In equations (\ref{eq_1}-\ref{eq_2}) and below velocity, time and distance are measured in units 
of $g_z t_s$, $\Omega_0^{-1}$ and $g_z t_s / \Omega_0$, respectively. The dimensionless epicyclic frequency 
squared is denoted by $\tilde \kappa^2 \equiv 2(2-q)$.

Equations (\ref{eq_1}-\ref{eq_2}) are derived up to the second order in the small ratios of the amplitudes 
of perturbations to the corresponding background quantities\footnote{
Note that the term '$\tau \nabla^2 W$' coming to the right-hand side of equation (\ref{eq_2}) from equations 
(\ref{eq_TVA}-\ref{eq_rho_g}) written for perturbations was excluded using the divergence of equation (\ref{eq_1})
up to the leading order in $\tau$.}.
Equations (\ref{eq_1}-\ref{eq_2}) are valid up to the linear order in the small dust fraction 
\begin{equation}
\label{dust_frac}
f \equiv \frac{\rho_p}{\rho_g} \ll 1
\end{equation} 
as the terms of higher order in $f$ has been omitted.
By the same reason, $W$ entering the non-linear term in RHS of equation (\ref{eq_1}) can be excluded according to the relation derived from the divergence of equation (\ref{eq_1}),
\begin{equation}
\label{eq_3}
W \approx 2 \nabla^{-2} \partial_x u_y,
\end{equation}
which is valid to the zeroth order in $f$.

\section{The linear problem}
\label{sec_linear}

\subsection{Dispersion equation}

Let perturbations be described by the vector of the state variables 
\begin{equation}
\label{chi}
\chi \equiv \{ u_x, u_y, u_z, \delta \}. 
\end{equation}
The particular solution for infinitesimal perturbations should be sought in the form of the Fourier harmonics
\begin{equation}
\label{chi_harmonic}
\chi^i = \hat \chi^i \exp (-{\rm i} \omega t + {\rm i} {\bf k}{\bf x}),
\end{equation}
where ${\bf k x} = k_x x + k_z z$.

The linearised equations (\ref{eq_1}-\ref{eq_2}) yield that $\hat \chi^i$ obey the following linear algebraic 
set of equations\footnote{In order to derive the RHS of equation (\ref{lin_eq_3}) ${\bf u}$ was assumed to be free 
of divergence, since the omitted terms $\propto f\tau$ go beyond the terminal velocity approximation, 
see the corresponding analysis in \citetalias{zhuravlev-2019} and recently in \citet{zhuravlev-2021}.}:
\begin{equation}
\label{lin_eq_1}
\omega k_z \hat u_x - \omega k_x \hat u_z = 2 {\rm i} k_z \hat u_y + {\rm i} \frac{f}{\tau} k_x \hat \delta,
\end{equation}

\begin{equation}
\label{lin_eq_2}
\omega k_z \hat u_y = - \frac{{\rm i\tilde \kappa^2}}{2} k_z \hat u_x,
\end{equation}

\begin{equation}
\label{lin_eq_3}
\omega k_x \hat u_y = 
\frac{{\rm i \tilde\kappa^2}}{2} k_z \hat u_z,
\end{equation}

\begin{equation}
\label{lin_eq_4}
-{\rm i} \omega \hat\delta = -2{\rm i} \tau k_x \hat u_y + {\rm i} (1-f) k_z \hat \delta,
\end{equation}
which gives the dispersion equation
\begin{equation}
\label{dispersion}
D_g^{+}(\omega,{\bf k}) \cdot D_g^{-}(\omega,{\bf k}) \cdot D_p(\omega,{\bf k}) = \epsilon({\bf k}),
\end{equation}
where
\begin{equation}
\label{D_g}
D_g^{\pm}(\omega,{\bf k}) \equiv \omega \mp  \omega_i, 
\end{equation}
\begin{equation}
\label{D_p}
D_p(\omega,{\bf k}) \equiv \omega - \omega_p,
\end{equation}
\begin{equation}
\label{epsilon}
\epsilon({\bf k}) \equiv f \, \tilde \kappa^2 \, \frac{k_x^2}{k^2} k_z
\end{equation}
with 
$k^2 \equiv k_x^2+k_z^2$,
$
\omega_p \equiv - k_z (1-f),
$
and
$\omega_i = \tilde \kappa \, k_z/k$.

As $\epsilon\to 0$, equation (\ref{dispersion}) splits into three separate dispersion equations explicitly 
given by the definitions (\ref{D_g}-\ref{D_p}), which describe two oppositely propagating IW and one SDW, 
see the details in \citetalias{zhuravlev-2019}.


Equations (\ref{lin_eq_1}-\ref{lin_eq_4}) reduce to equations (31-34) of \citetalias{zhuravlev-2019} in the 
vicinity of the linear resonance, or alternatively, the mode crossing between SDW and IW, 
where the non-resonant correction $\propto f \hat\delta$ to the dynamics of SDW in RHS of equation (\ref{lin_eq_4})
becomes small compared to the coupling term in RHS of the dispersion equation (\ref{dispersion}).
In the opposite case, when considering the solution of equation (\ref{dispersion}) sufficiently far
from the mode crossing, although staying within the limit of the small dust fraction, this non-resonant correction
should be retained along with the (non-resonant) correction due to the coupling term, $\epsilon$. 
Below, such a linear solution will be referred to as the non-resonant one in contrast to the resonant solution considered 
in Section 3.5 of \citetalias{zhuravlev-2019}. It turns out that the resonant solution describes the growing modes of DSI with growth rates 
$\propto \epsilon^{1/2}$, while the non-resonant solution describes the neutral modes with real corrections to the frequencies $\propto \epsilon$, 
see the next Section.

\subsection{Approximate non-resonant solution in the limit of small $f$}
\label{sec_non_res_sol}

The solution of equation (\ref{dispersion}) can be sought in the form
\begin{equation}
\label{delta_pm}
\omega = \pm \omega_i + \Delta_i^\pm
\end{equation}
as well as
\begin{equation}
\label{delta_p}
\omega = \omega_p + \Delta_p,
\end{equation}
where it is assumed that the corrections are small,
\begin{equation}
\label{iw_appr}
|\Delta_i^\pm| \ll \omega_i, 
\end{equation}
\begin{equation}
\label{sdw_appr}
|\Delta_p| \ll |\omega_p|. 
\end{equation}
In this case, equation (\ref{dispersion}) can be reduced to the corresponding quadratic equations with respect to $\Delta_p$ and $\Delta_{\pm}$.
They read
\begin{equation}
\label{quadr_1}
(\omega_p^2 - \omega_i^2 + 2\omega_p \Delta_p) \Delta_p = \epsilon
\end{equation}
and
\begin{equation}
\label{quadr_2}
\pm 2\omega_i \Delta_\pm (\pm\omega_i-\omega_p + \Delta_\pm) = \epsilon.
\end{equation}

In the limit $\pm \omega_i \to \omega_p$ equations (\ref{quadr_1}-\ref{quadr_2}) have the same resonant solution $\propto \epsilon^{1/2}$, 
which describes DSI. In the opposite limit, when the solution to equations (\ref{quadr_1}-\ref{quadr_2}) is considered far from the mode crossing 
between SDW and IW, $\pm \omega_i=\omega_p$, so that
\begin{equation}
\label{iw_not_res}
2\epsilon \ll \omega_i (\omega_i \mp\omega_p)^2
\end{equation}
for $\Delta_\pm$ and
\begin{equation}
\label{sdw_not_res}
8\epsilon |\omega_p| \ll (\omega_i^2-\omega_p^2)^2 
\end{equation}
for $\Delta_p$, respectively, 
one obtains to leading order in $f$
\begin{equation}
\label{correct_i_pm}
\Delta_i^\pm \approx \frac{\epsilon}{2(\omega_i \mp \omega_p)\,\omega_i}
\end{equation}
and
\begin{equation}
\label{correct_p_pm}
\Delta_p \approx \frac{\epsilon}{\omega_p^2 - \omega_i^2},
\end{equation}
where it can be assumed that $\omega_p=-k_z$.

The inequalities (\ref{iw_appr}-\ref{sdw_appr}) and (\ref{iw_not_res}-\ref{sdw_not_res}) 
will be used in Section \ref{sec_t_e} to evaluate the bounds of the analytical model. 

In this way, one finds the approximate solutions of equation (\ref{dispersion}) taking into account 
the non-zero coupling between SDW and IW provided that it occurs sufficiently far from the mode crossing.
In what follows, it is assumed that $k_z>0$.
For the given wavenumber, ${\bf k}$, explicitly,

\begin{equation}
\label{sdw_freq}
\omega_{({\bf k})[1]} \approx \omega_p + f \, \tilde\kappa^2\, \frac{k_x^2}{k_z k^2} \left ( 1 - \frac{\tilde\kappa^2}{k^2}  \right )^{-1}
\end{equation}
corresponding to the slightly modified counterpart of SDW, and
\begin{equation}
\label{iwpm_freq}
\omega_{({\bf k})[2],[3]} \approx \mp \omega_i \mp \frac{f}{2} \, \tilde\kappa \, \frac{k_x^2}{k_z k} \left ( 1 \mp \frac{\tilde\kappa}{k} \right )^{-1}
\end{equation}
corresponding to the slightly modified counterparts of IW$^{-}$ and IW$^{+}$ propagating, respectively, in the same and the opposite sense as SDW. 
Hereafter, the modes with frequencies (\ref{sdw_freq}) and (\ref{iwpm_freq})
will be referred to as SDW and IW$^\mp$, respectively.
Note that, here and below, the index $1,2,3$ in square brackets in the left-hand side of equations (\ref{sdw_freq}-\ref{iwpm_freq}) stands, respectively, for SDW, IW$^{-}$ and IW$^{+}$.

\section{Three-wave resonance}
\label{sec_res}

\begin{figure}
\begin{center}
\includegraphics[width=8cm,angle=0]{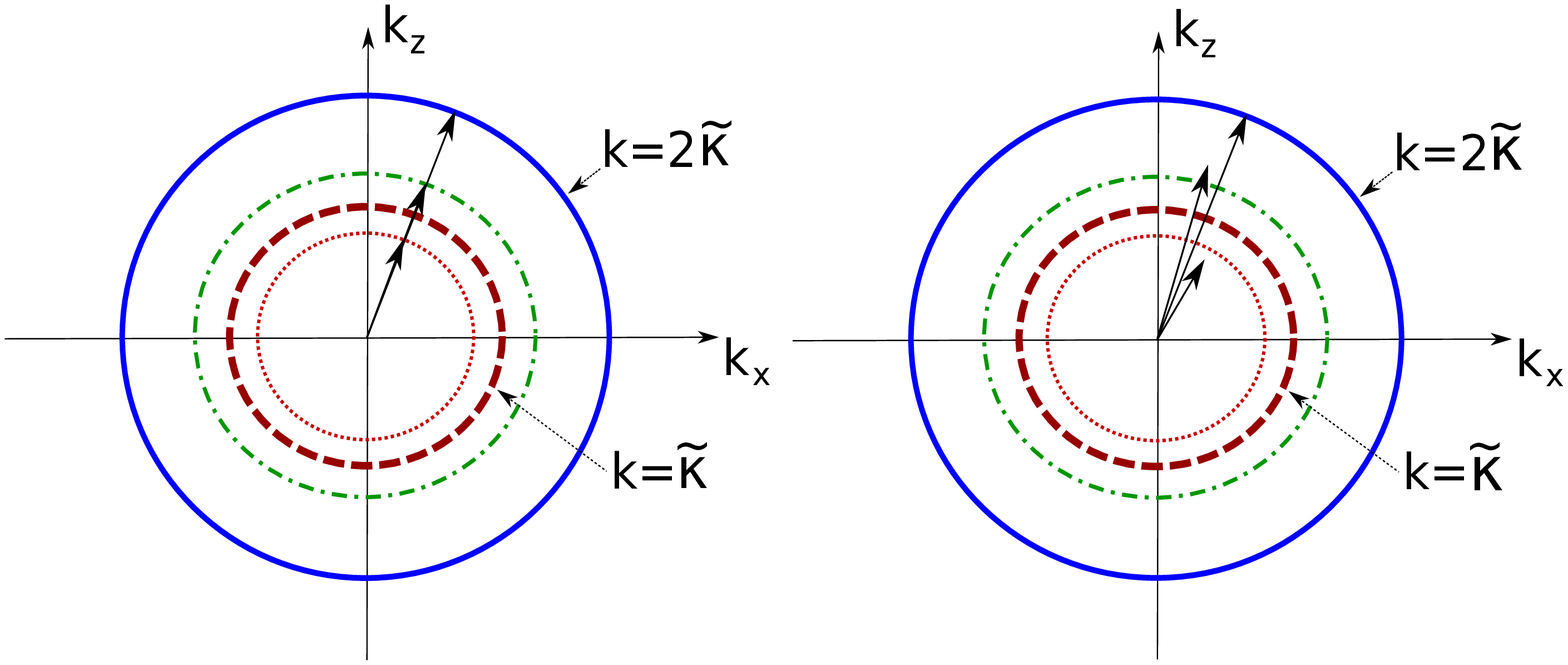}
\end{center}
\caption{The resonant waves on the plane of wavenumbers. Solid (blue), dashed (brown), dotted (red) and 
dot-dashed (green) lines show the absolute wavenumber of, respectively, SDW$_r$, DSI at the mode crossing between SDW and IW, 
IW$^\prime_r$ and IW$^{\prime\prime}_r$. 
The black arrows represent the particular waves in the case $f\to 0$ (left panel) and in the case $f>0$ (right panel). 
} \label{fig}
\end{figure}

Let the resonant triad contain one negative energy SDW, which will be additionally marked with an index `r' below.
In this case, in order for the resonant triad to satisfy the conditions of explosive instability, 
the two remaining waves should have positive energies and propagate in the same sense as SDW. 
Therefore, along with SDW the explosive resonant triad should contain two IW$^{-}$. 
These waves will be denoted as IW$^\prime_r$ and IW$^{\prime\prime}_r$ hereafter.
Thus, the three-wave resonance considered in this work has the following form
\begin{equation}
\label{triad}
{\rm SDW}_r \to {\rm IW}^\prime_r + {\rm IW}^{\prime\prime}_r.
\end{equation}

The waves entering (\ref{triad}) must satisfy the following condition of three-wave resonance
\begin{equation}
\label{three_wave_gen}
\omega_{({\bf k}^\bullet)[1]} = \omega_{({\bf k}^\prime)[2]} + \omega_{({\bf k}^{\prime\prime})[2]},
\end{equation}
where
\begin{equation}
\label{three_k_gen}
{\bf k}^\bullet = {\bf k}^\prime + {\bf k}^{\prime\prime}
\end{equation}
are the wavenumbers to be determined. This condition comes from the requirement that each of three modes matches both spatial and temporal periodicities 
of the driving force arising due to interaction between the other two modes, see e.g. \citet{kadomtsev-1971} and \citet{craik_book}.


It is not difficult to consider the existence of resonant triads (\ref{triad}) on the plane of 
wavenumbers in the limit of negligible $f\to 0$.
Let $\theta^\bullet$, $\theta^\prime$ and $\theta^{\prime\prime}$ be the angles of ${\bf k}^\bullet$, ${\bf k}^\prime$ and ${\bf k}^{\prime\prime}$, 
respectively, with respect to the radial direction. 
Equations (\ref{three_wave_gen}) and (\ref{three_k_gen}) yield in this case
\begin{equation}
\label{gen_res_sys}
\begin{aligned}
&\tilde \kappa (\sin \theta^\prime + \sin \theta^{\prime\prime}) = k^\bullet \sin \theta^\bullet,\\
&k^\prime \sin \theta^\prime + k^{\prime\prime} \sin \theta^{\prime\prime} = k \sin \theta^\bullet \\
&k^\prime \cos \theta^\prime + k^{\prime\prime} \cos \theta^{\prime\prime} = k \cos \theta^\bullet.
\end{aligned}
\end{equation}
The set of equations (\ref{gen_res_sys}) has the following solution provided that the wave angles are known
\begin{equation}
\label{gen_res_sol}
\begin{aligned}
&k^\bullet = 2 \tilde \kappa \cos \left ( \frac{\beta^\prime + \beta^{\prime\prime}}{2} \right ), \\
&k^{\prime} = k^\bullet \, \frac{\sin (\beta^{\prime\prime})}{\sin (\beta^\prime + \beta^{\prime\prime})}, \\
&k^{\prime\prime} = k^\bullet \, \frac{\sin (\beta^{\prime})}{\sin(\beta^\prime + \beta^{\prime\prime})},
\end{aligned}
\end{equation}
where $k^\bullet$, $k^\prime$ and $k^{\prime\prime}$ stand for the wavenumbers of SDW$_r$, IW$^\prime_r$ and IW$^{\prime\prime}_r$, respectively, 
while 
$
\beta^\prime \equiv \theta^\prime - \theta^\bullet
$ 
and 
$
\beta^{\prime\prime} \equiv \theta^\bullet - \theta^{\prime\prime}
$ 
are the angles between the resonant waves.
In order to satisfy the general conditions of explosive instability, 
the resonant triad (\ref{gen_res_sol}) is obtained assuming that $0 < \theta^{\prime\prime} < \theta^\bullet < \theta^{\prime} < \pi$. 
It can be seen that the existence of this triad is quite general.
It follows that the wavenumber of SDW$_r$ covers the range $(0,2\tilde\kappa)$, while the wavenumbers of the two other resonant waves
take values from $0$ up to the wavenumber of SDW$_r$ depending on the ratio between the angles $\beta^\prime$ and $\beta^{\prime\prime}$. At the same time, 
as soon as the angles $\beta^\prime$ and $\beta^{\prime\prime}$ are specified, the resonant triad exists regardless of the direction of 
wave propagation.

The solution (\ref{gen_res_sol}) should be used to find the small corrections to ${\bf k}^\bullet$, ${\bf k}^\prime$ and ${\bf k}^{\prime\prime}$
caused by the non-zero dust fraction. These are determined by an equation (\ref{three_wave_gen}) combined with the approximate frequencies 
(\ref{sdw_freq}-\ref{iwpm_freq}). 
This exercise is straightforward, however, it seems to entail formidable algebraic calculations of interaction between the resonant waves, 
see Section \ref{sec_coupling}.
As the present study aims to treat the interactions within the resonant triad by the analytical means, it is confined below to
a simple particular case
of collinear waves, $\beta^{\prime},  \beta^{\prime\prime} \to 0$.
For the special choice $a \beta^\prime = (1-a) \beta^{\prime\prime}$, the solution (\ref{gen_res_sol}) yields  
\begin{equation}
\label{three_wave_sol}
{\bf k}^\prime = a {\bf k}^\bullet, \quad {\bf k}^{\prime\prime} = (1-a) {\bf k}^\bullet \quad \mbox{and} \quad 
|{\bf k}^\bullet|=2\, \tilde\kappa,
\end{equation}
where the dimensionless free parameter $a$ is assumed to be enclosed in the range $0<a<1/2$.
Thus, ${\rm IW}^\prime_r$ and ${\rm IW}^{\prime\prime}_r$ 
are both collinear to ${\rm SDW}_r$ but 
$k > k^{\prime\prime} > k^\prime$, see the left panel in Figure \ref{fig}.
Note that as $a\to 1/2$, ${\bf k}^\prime \to {\bf k}^{\prime\prime} \to {\bf k}_{\rm DSI}$,
where $k_{\rm DSI} = \tilde \kappa$, which means that ${\rm IW}^\prime_r$ and ${\rm IW}^{\prime\prime}_r$ 
approach each other at the mode crossing with another SDW defined by the equality $-\omega_i = \omega_p$. 
In this study $a$ cannot take value too close to $1/2$ for the sake of possibility of the analytical treatment, 
see Section \ref{sec_non_res_sol}.
Note that for waves from the solution (\ref{gen_res_sol}) as well as (\ref{three_wave_sol}) 
the requirement $\tau \ll 1$ is sufficient to satisfy the conditions (\ref{TVA_1}-\ref{TVA_2}) of 
the terminal velocity approximation.

It is not difficult to see that interaction between the waves of the resonant triad vanishes for $f\to 0$, since 
in this limit the operator ${\bf u} \cdot \nabla$ acting on any field collinear to ${\bf u}$ results with 
the zero value for the divergence-free ${\bf u}$, 
what implies that non-linear terms in equations (\ref{eq_1}-\ref{eq_2}) vanish. 

As soon as the dust fraction is not negligible, $f>0$, 
${\rm SDW}_r$  
should interact with slightly non-collinear ${\rm IW}^\prime_r$ and ${\rm IW}^{\prime\prime}_r$ 
having wavenumbers ${\bf k}^\prime + \Delta {\bf k}$ and ${\bf k}^{\prime\prime} - \Delta {\bf k}$, 
respectively, see the right panel in Figure \ref{fig}.
To keep the notations simple, everywhere below it is assumed that 
${\bf k}^\prime$ and ${\bf k}^{\prime\prime}$ contain the small correction $\Delta {\bf k}$,
which is proportional to $f$. This correction is derived from equation (\ref{three_wave_gen}) combined with equations
(\ref{sdw_freq}), (\ref{iwpm_freq}) and (\ref{three_wave_sol}) to leading order in the small dust fraction.
The corresponding small differences to the resonant frequencies 
\begin{equation}
\label{three_wave_gen_delta}
\Delta \omega_{\rm SDW} = \Delta \omega_{\rm IW^\prime} + \Delta \omega_{\rm IW^{\prime\prime}}
\end{equation}
are explicitly 
\begin{equation}
\label{Delta_freq_SDW}
\Delta \omega_{\rm SDW} = f k_z^\bullet \left ( 1 + \frac{{k_x^\bullet}^2}{3{k_z^\bullet}^2} \right ),
\end{equation}
\begin{equation}
\label{Delta_freq_IW}
\Delta \omega_{\rm IW^\prime} = f \frac{a}{1-2a} \frac{{k_x^\bullet}^2}{2 k_z^\bullet} - 
\frac{k_x^\bullet}{8a \tilde \kappa^2} ( k_x^\bullet \Delta k_z - k_z^\bullet \Delta k_x )
\end{equation}
and $\Delta \omega_{\rm IW^{\prime\prime}}$ is given by equation (\ref{Delta_freq_IW}) with the replacement $a \to 1-a$.
Equation (\ref{three_wave_gen_delta}) leads to the following condition on $\Delta {\bf k}$
specified by its own cross product with ${\bf k}^\bullet$:
\begin{equation}
\label{three_wave_cond}
k_x^\bullet \Delta k_z - k_z^\bullet \Delta k_x = 
f \, \tilde\kappa^2 \, \frac{8 a(1-a)}{2a-1} \left ( \frac{k_z^\bullet}{k_x^\bullet} + \frac{5}{6} \frac{k_x^\bullet}{k_z^\bullet} \right ).
\end{equation}

In order to obtain the frequencies of the resonant triad for $f>0$, 
one should use equations (\ref{sdw_freq}) and (\ref{iwpm_freq})
in combination with the conditions (\ref{three_wave_sol}) and (\ref{three_wave_cond}), 
see Appendix \ref{freqs} for the resulting expressions.
With the triad frequencies at hand, the linear equations (\ref{lin_eq_1}-\ref{lin_eq_4}) 
provide the eigen-vectors, $\hat \chi$, corresponding to SDW$_r$, IW$^{\prime}_r$ and IW$^{\prime\prime}_r$,
see Appendix \ref{basis} for the resulting expressions.

Complementary eigen-frequencies and eigen-vectors 
provided, respectively, by equations (\ref{sdw_freq} - \ref{iwpm_freq}) and  
(\ref{lin_eq_1}-\ref{lin_eq_4}) at each of the triad wavenumbers
are obtained in a similar way, see Appendixes \ref{freqs} and \ref{basis}.
For the triad wavenumbers, the corresponding sets of eigen-vectors
construct the three different bases necessary to obtain the coupling coefficients of wave-wave interaction, 
see the next Section.

\section{Interaction of waves}
\label{sec_coupling}

The finite-amplitude perturbations can be considered in the form of the spatial Fourier harmonics 
$$
\chi^i = \psi_{({\bf k})}^i \exp({\rm i} {\bf k x})
$$
at the arbitrary wavenumber ${\bf k}$.

It follows from equations (\ref{eq_1}-\ref{eq_2}) that $\psi_{{\bf k}\,i}$ satisfies an equation
\begin{equation}
\label{eq_psi}
\partial_t \psi_{({\bf k})}^i = \sum_{j=1}^{4} L_{({\bf k})j}^i \psi_{({\bf k})j} + 
\frac{1}{2\pi} \sum_{j,k=1}^{4} \int \limits_{-\infty}^{+\infty} N_{({\bf k}, {\bf l}) jk}^i \, \psi_{({\bf k}- {\bf l})}^j \ \psi_{({\bf l})}^k \,
d {\bf l},
\end{equation}
where $L_{({\bf k})j}^i$ specifies the linear dynamics of perturbations, while $N_{({\bf k}, {\bf l}) jk}^i$ 
describes their quadratic interaction. 
Explicitly, 
\begin{equation}
\label{L_ij}
L_{({\bf k}) j}^i = \left ( 
\begin{array}{cccc}
0 & 2\frac{k_z^2}{k^2} & 0 & \frac{f}{\tau}\frac{k_x k_z}{k^2} \\
-\frac{\tilde \kappa^2}{2} & 0 & 0 & 0 \\
0 & -2\frac{k_x k_z}{k^2} & 0 & -\frac{f}{\tau} \frac{k_x^2}{k^2} \\
0 & -2{\rm i} \tau k_x & 0 & {\rm i} k_z (1-f)
\end{array}
\right ),
\end{equation}
\begin{equation}
\label{N_1}
N_{({\bf k}, {\bf l}) jk}^1 = \left ( 
\begin{array}{cccc}
-{\rm i} l_{x} & 0 & 0 & 0 \\
0 & 0 & 0 & 0 \\
-{\rm i} l_{z} & 0 & 0 & 0 \\
0 & 2 f \frac{l_x^2}{l^2} & 0 & 0
\end{array}
\right ),
\end{equation}

\begin{equation}
\label{N_2}
N_{({\bf k}, {\bf l}) jk}^2 = \left ( 
\begin{array}{cccc}
0 & -{\rm i} l_{x} & 0 & 0 \\
0 & 0 & 0 & 0 \\
0 & -{\rm i} l_{z} & 0 & 0 \\
0 & 0 & 0 & 0
\end{array}
\right ),
\end{equation}

\begin{equation}
\label{N_3}
N_{({\bf k}, {\bf l}) jk}^3 = \left ( 
\begin{array}{cccc}
0 & 0 & -{\rm i} l_{x} & 0 \\
0 & 0 & 0 & 0 \\
0 & 0 & -{\rm i} l_{z} & 0 \\
0 & 2 f \frac{l_x l_z}{l^2} & 0 & 0
\end{array}
\right ),
\end{equation}

\begin{equation}
\label{N_4}
N_{({\bf k}, {\bf l}) jk}^4 = \left ( 
\begin{array}{cccc}
-\tau (k_x - l_x) l_x & 0 & - \tau (k_z-l_z) l_x & 0 \\
0 & 0 & 0 & 0 \\
-\tau (k_x-l_x) l_z & 0 & - \tau (k_z-l_z) l_z & 0 \\
-{\rm i} k_x & 0 & -{\rm i} k_z & {\rm i} f k_z
\end{array}
\right ).
\end{equation}

The general solution of equation (\ref{eq_psi}) can be sought as the combination of linear modes
\begin{equation}
\label{non_lin_sol}
\psi_{({\bf k})}^i = \sum_{s=1}^{3} A_{({\bf k}) [s]} (t) \phi_{({\bf k}) [s]}^i \exp({-\rm i} \omega_{({\bf k})[s]} t ), 
\end{equation}
with amplitudes $A_{({\bf k}) [s]}$ evolving due to interaction of modes at different wavenumbers.
Note that, in accordance with the notations in equations (\ref{sdw_freq}-\ref{iwpm_freq}), 
the values of the new index in the square brackets, $s=1,2,3$, correspond to SDW, IW$^-$ and IW$^+$, respectively.

Written in this way, vectors $\phi_{({\bf k})[s]}^i$ are equivalent to $\hat \chi^i$ being the solution
of the linear system (\ref{lin_eq_1}-\ref{lin_eq_4}). Consequently, they are the eigen-vectors of $L_{({\bf k})j}^i$,
\begin{equation}
\label{eigen_L}
-{\rm i} \omega_{({\bf k})[s]} \phi_{({\bf k}) [s]}^i = \sum_{j=1}^4 L_{({\bf k})j}^i \phi_{({\bf k})[s]}^j.
\end{equation}
These eigen-vectors construct basis in the linear space of vectors $\psi$ corresponding to the particular wavenumber.
Note that equation (\ref{eigen_L}) is satisfied separately for each $s=1,2,3$.

Equation (\ref{eq_psi}) comes to
\begin{equation}
\label{eq_psi_1}
\begin{aligned}
\sum_{s=1}^{3} \partial_t A_{({\bf k})[s]} \phi_{({\bf k}) [s]}^i \exp(-{\rm i} \omega_{({\bf k})[s]} t)  = 
\frac{1}{2\pi} \sum_{p,q=1}^{3} \sum_{j,k=1}^{4}  \int \limits_{-\infty}^{+\infty} N_{({\bf k},{\bf l})jk}^i \\ 
A_{({\bf k}-{\bf l})[p]} A_{({\bf l})[q]} \phi_{({\bf k}-{\bf l}) [p]}^j \phi_{({\bf l}) [q]}^k 
\exp(-{\rm i} \omega_{({\bf k}-{\bf l})[p]} t -{\rm i} \omega_{({\bf l})[q]} t) d{\bf l}. 
\end{aligned}
\end{equation}

The linear space of vectors represented by equation (\ref{non_lin_sol}) 
can be normalised according to the following inner product
\begin{equation}
\label{inner_prod}
(\psi_1,\psi_2) = \sum_{i=1}^{4} \psi_{1 ({\bf k})}^i \psi_{2 ({\bf k})}^{i\,*},
\end{equation}
where the asterisk denotes complex conjugation. 
The rule (\ref{inner_prod}) allows one to introduce the dual basis
\begin{equation}
\label{rule_dual}
( \phi_{[p]}, \tilde \phi_{[q]} ) = \delta_{pq},
\end{equation}
where $\delta_{pq}$ is the Kronecker delta.

Having the vectors of dual basis at hand, one finds its inner product with equation 
(\ref{eq_psi_1}) in the following form\footnote{The duality of $\tilde \phi$ guarantees the results be independent of 
the specific form of the inner product.}
\begin{equation}
\label{eq_psi_fin}
\begin{aligned}
\partial_t A_{({\bf k})[\tilde s]} \exp(-{\rm i} \omega_{({\bf k})[\tilde s]} t)  = 
\frac{1}{2\pi} \sum_{p,q=1}^{3}  \int \limits_{-\infty}^{+\infty}  Q_{({\bf k},{\bf l})[\tilde s][p][q]} \\ 
A_{({\bf k}-{\bf l})[p]} A_{({\bf l})[q]} 
\exp(-{\rm i} \omega_{({\bf k}-{\bf l})[p]} t -{\rm i} \omega_{({\bf l})[q]} t) d{\bf l}, 
\end{aligned}
\end{equation}
where
\begin{equation}
\label{Q_coeffs}
Q_{({\bf k},{\bf l})[\tilde s][p][q]} \equiv \sum_{i,j,k=1}^{4}  N_{({\bf k},{\bf l}) jk}^i \tilde \phi_{({\bf k}) [\tilde s]}^{i\,*} 
\phi_{({\bf k}-{\bf l}) [p]}^j \phi_{({\bf l}) [q]}^k.
\end{equation}

From now on, it is assumed that the modes are excited only at the wavenumbers of the three-wave resonance proposed
in Section \ref{sec_res}.
The eigen-vectors $\phi_{({\bf k}^\bullet) [p]}^i$, $\phi_{({\bf k}^\prime) [p]}^i$ and $\phi_{({\bf k}^{\prime\prime}) [p]}^i$ 
obtained at these wavenumbers can be found in Appendix \ref{basis}. 
Accordingly, the necessary vectors of dual basis are given in Appendix \ref{dual_basis}. 
Equation (\ref{eq_psi_fin}) yields the following set of evolutionary equations for amplitudes
of the resonant triad.
First, for SDW$_r$
\begin{equation}
\label{eq_A_1}
\partial_t A_{({\bf k}^\bullet)[1]} = Q_1 A_{({\bf k}^\prime)[2]} A_{({\bf k}^{\prime\prime})[2]},
\end{equation}
where
\begin{equation}
\label{C_1}
\begin{aligned}
Q_1 = \frac{1}{2\pi} \sum_{i,j,k=1}^{4} & \left [ N_{({\bf k}^\bullet,{\bf k}^\prime) jk}^i \phi_{({\bf k}^{\prime\prime}) [2]}^j 
\phi_{({\bf k}^\prime) [2]}^k +  \right . \\
&\left . N_{({\bf k}^\bullet,{\bf k}^{\prime\prime}) jk}^i \phi_{({\bf k}^{\prime}) [2]}^j \phi_{({\bf k}^{\prime\prime}) [2]}^k 
\right ] \tilde \phi_{({\bf k}^\bullet) [1]}^{i\, *}.
\end{aligned}
\end{equation}
Next, for IW$^\prime_r$
\begin{equation}
\label{eq_A_2}
\partial_t A_{({\bf k}^\prime)[2]} = Q_2 A_{({\bf k}^\bullet)[1]} A_{({\bf k}^{\prime\prime})[2]},
\end{equation}
where
\begin{equation}
\label{C_2}
\begin{aligned}
Q_2 = \frac{1}{2\pi} \sum_{i,j,k=1}^{4} & \left [ N_{({\bf k}^\prime,{\bf k}^\bullet) jk}^i \phi_{({\bf k}^{\prime\prime}) [2]}^{j\, *} 
\phi_{({\bf k}^\bullet) [1]}^k 
+ \right . \\ 
& \left .
N_{({\bf k}^\prime,-{\bf k}^{\prime\prime}) jk}^i \phi_{({\bf k}^\bullet) [1]}^j 
\phi_{({\bf k}^{\prime\prime}) [2]}^{k\, *} \right ]  \tilde \phi_{({\bf k}^\prime) [2]}^{i\, *}.
\end{aligned}
\end{equation}

Last, for IW$^{\prime\prime}_r$
\begin{equation}
\label{eq_A_3}
\partial_t A_{({\bf k}^{\prime\prime})[2]} = Q_3 A_{({\bf k}^\bullet)[1]} A_{({\bf k}^{\prime})[2]},
\end{equation}
where
\begin{equation}
\label{C_3}
\begin{aligned}
Q_3 = \frac{1}{2\pi} \sum_{i,j,k=1}^{4} & \left [ 
N_{({\bf k}^{\prime\prime},{\bf k}^\bullet) jk}^i \phi_{({\bf k}^\prime) [2]}^{j\, *} 
\phi_{({\bf k}^\bullet) [1]}^k 
+ \right . \\
& \left .
N_{({\bf k}^{\prime\prime},-{\bf k}^{\prime}) jk}^i \phi_{({\bf k}^\bullet) [1]}^j 
\phi_{({\bf k}^{\prime}) [2]}^{k\, *}  \right ]  \tilde \phi_{({\bf k}^{\prime\prime}) [2]}^{i\, *}.
\end{aligned}
\end{equation}


In order to shorten the notations further on, the following changes are made 
$$
\begin{aligned}
& A_{({\bf k}^\bullet)[1]} \to A_1, \\ 
& A_{({\bf k}^\prime)[2]} \to A_2, \\ 
& A_{({\bf k}^{\prime\prime})[2]} \to A_3,
\end{aligned}
$$
which end up in equations for three-wave resonance
\begin{equation}
\label{system_non_lin}
\begin{aligned}
\partial_t A_1 = Q_1 A_2 A_3,\\
\partial_t A_2 = Q_2 A_1 A_3,\\
\partial_t A_3 = Q_3 A_1 A_2.
\end{aligned}
\end{equation}
Note that $A_1$ is the relative dust density perturbation, $\hat\delta$,
in SDW$_r$, while $A_2$ and $A_3$ are the radial projections of the velocity perturbation, $\hat u_x$, 
in IW$_r^\prime$ and IW$_r^{\prime\prime}$, respectively.

Derivation of the coupling coefficients is straightforward and its details can be found in Appendix \ref{coupl_coeffs}.
Explicitly, 
\begin{equation}
\label{Q_1}
Q_1 = - \frac{16}{3\pi} f\tau \tilde \kappa^4 \, \frac{a(1-a) \, ( 20\,\tilde\kappa^2 + k_z^2 ) }{ (2a-1)^2 k_z^4 },
\end{equation}
\begin{equation}
\label{Q_2}
Q_2 = Q_3 = - \frac{1}{3\pi} \frac{f^2}{\tau} \frac{ a(1-a) \, ( 20\, \tilde\kappa^2 + k_z^2 ) }{ (2a-1)^2 k_z^2 }.
\end{equation}
Note that the superscript $\bullet$ after $k_x$ and $k_z$ is omitted in equations (\ref{Q_1}), (\ref{Q_2}) and everywhere below. 

\begin{figure}
\begin{center}
\includegraphics[width=6cm,angle=0]{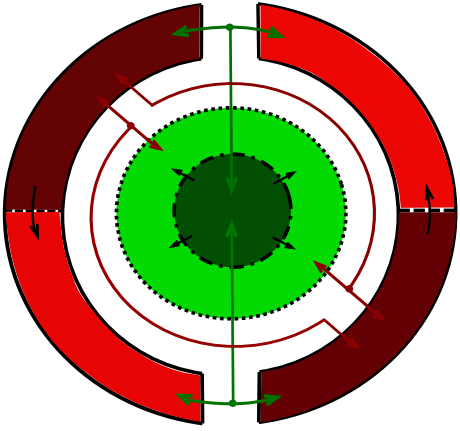}
\end{center}
\caption{A scheme of interaction between the resonant waves. The two (red) semi-circular domains framed by solid lines represent IW$_r^\prime$ 
and IW$_r^{\prime\prime}$, while the (green) circular central domain framed by dotted line represents SDW$_r$. 
Each wave consists of the primary part (dark-coloured) and the secondary part (light-coloured). The latter is always induced by the former. 
The secondary parts of both IW$_r$, which are perturbations of the dust density $\sim \tau$, 
arise due to the aerodynamical concentration of particles in gas eddies being the primary parts of these waves. This process
is shown by the black arrows crossing the dashed lines inside the domains of IW$_r$'s. 
The secondary part of SDW$_r$, which is perturbation of gas velocity $\sim f$, arises due to the dust back reaction on gas.
This process is shown by the black arrows crossing the dot-dashed circle inside the domain of SDW$_r$.
Interactions between IW$_r$'s are shown by the green double-sided arrows connecting their corresponding primary and secondary parts.
As shown on the scheme, this process generates the primary perturbations of the dust density in SDW$_r$. 
Interactions between either of IW$_r$ and SDW$_r$ are shown by the red double-sided arrows connecting their primary and secondary parts, respectively. 
As shown on the scheme, this process generates the primary perturbations of the gas velocity in the other IW$_r$.  
} \label{fig_2}
\end{figure}

Examination of equations (\ref{C_1}), (\ref{C_2}), (\ref{C_3}) along with the structure of eigen-vectors, see
Appendixes \ref{basis} and \ref{dual_basis}, reveals the general physics of the interaction between the resonant
waves, which is illustrated by the scheme in Figure \ref{fig_2}.
Namely, the first equation of (\ref{system_non_lin}) drives the perturbation of the dust density in SDW$_r$
caused mainly by the second non-linear term in equation (\ref{eq_2}) being the product of the velocity perturbation
of one IW$_r$ and the dust density perturbation of the other IW$_r$ induced aerodynamically by the gradient of its own
perturbation of pressure.
The second and the third equations of (\ref{system_non_lin}) drive perturbations of velocity in either IW$_r$
caused mainly by the advection term in the left-hand side of equation (\ref{eq_1}) being the product 
of the velocity perturbation of the other IW$_r$ and the velocity perturbation of SDW$_r$ induced by its own 
perturbation of the dust density via the dust back reaction on the gas.


There is a caution concerning the dependence of the coupling coefficients on the dust fraction.
Indeed, an extra order of $f$ in all $Q_{1,2,3}$ comes from the geometry of the triad, i.e. its 
collinearity in the limit $f\to 0$. The coupling coefficients driving IW$^{\prime,\prime\prime}_r$ 
of some other triad (\ref{triad}) consisting of the non-collinear waves in the zeroth order in $f$ 
should be as small as $Q_{2,3} \sim f$, while SDW$_r$ should be driven already 
in the limit of negligible $f\to 0$\footnote{The latter, of course, does not mean that the dynamical loop shown in Figure \ref{fig_2}
persists for non-collinear triad in the limit $f\to 0$.}.


In this work, perturbation of velocity and therefore the amplitudes of IW$_r$ are measured in units of 
the dust settling velocity. In general, this is a physically justified choice, since the restriction $u \lesssim 1$ 
implies that the leading non-linear terms, $({\bf u} \cdot \nabla) {\bf u}$ in equation (\ref{eq_1}) and 
$\nabla \cdot (\delta {\bf u})$ in equation (\ref{eq_2}), are weaker than the linear terms in these equations at  $t_{ev} \sim \Omega_0^{-1}$ 
and $l_{ev} \sim g_z t_s / \Omega_0$ corresponding, respectively, to time- and length-scales of the resonant triad. 
Therefore, $u \lesssim 1$ corresponds to a weakly non-linear regime of perturbation dynamics. 
However, this restriction seems to be excessive in the particular case of perturbations, which construct SDW$_r$, IW$_r^\prime$ and IW$_r^{\prime\prime}$. 
That is it by the following reasons. First, the leading non-linear terms vanish for a single incompressible linear mode 
due to $\nabla \cdot {\bf u} = 0$, while the
non-vanishing terms are smaller at least by factor $f\ll 1$, see the last terms in RHS of equations (\ref{eq_1}) and (\ref{eq_2}).  
Consequently, IW$_r^\prime$ or IW$_r^{\prime\prime}$ alone can safely propagate in a mixture with amplitudes $u > 1$. This is even more true for
SDW$_r$, as its perturbation velocity induced by the dust back reaction on gas $\sim f$. 
Second, interaction between the resonant modes is weakened by the factors either $\sim \tau $ or $\sim f$ because it proceeds, respectively, 
either via aerodynamical concentration of dust or via the dust back reaction on gas, see Figure \ref{fig_2}. Moreover, an additional factor 
of $f$ weakening their interaction comes from the collinearity of the particular resonant triad considered in this work, see the caution 
made here above. In this situation, it is plausible to measure velocity perturbations in the usual units of sound speed, 
i.e. $c_s \equiv g_z / \Omega_0$, which is larger than the settling rate by factor $\tau^{-1}$.

The change to the units of $c_s$ leads to the replacement $Q_1 \to Q_1 / \tau^2$ in the first equation of (\ref{system_non_lin}).
That is, all coupling coefficients diverge as $\tau \to 0$ implying that 
the resonant interaction of modes becomes infinitely strong for small particles 
in spite of the vanishing settling as well as aerodynamic clumping. 
However, as $\tau\to 0$, the dimensional wavelengths of resonant waves, $\sim g_z t_s / \Omega_0 \sim \tau$, also vanish.
Thus, the corresponding unbounded amplification of resonant interaction is associated with 
the gradient standing in the leading non-linear terms 
of equations (\ref{eq_1}-\ref{eq_2}) and working at the vanishing scale of the three-wave resonance. 
Dissipative forces existing in a real disc should suppress the interaction of waves at scales smaller than
some threshold scale. The corresponding threshold value of $\tau \equiv \tau_\nu$ is estimated below in 
Section \ref{sec_Keplerian}. Note that the units of $c_s$ for the amplitudes of velocity perturbations will be used in Section \ref{sec_lower}
and there below.

Equations (\ref{Q_1}) and (\ref{Q_2}) show that $Q_{1,2,3}$ are negative-definite for the accepted values of $a$.
In this case, the following replacement
\begin{equation}
\label{repl_A}
\begin{aligned}
A_1 \to -\frac{\tilde A_1}{\sqrt{-Q_2} \sqrt{-Q_3}}, \\
A_2 \to -\frac{\tilde A_2}{\sqrt{-Q_1} \sqrt{-Q_3}}, \\
A_3 \to -\frac{\tilde A_3}{\sqrt{-Q_1} \sqrt{-Q_2}}
\end{aligned}
\end{equation}
leads to the set of equations of the standard type 
\begin{equation}
\label{system_non_lin_tilde}
\begin{aligned}
\partial_t \tilde A_1 = \tilde A_2 \tilde A_3,\\
\partial_t \tilde A_2 = \tilde A_1 \tilde A_3,\\
\partial_t \tilde A_3 = \tilde A_1 \tilde A_2.
\end{aligned}
\end{equation}

The solution of equations (\ref{system_non_lin_tilde}) can be found in terms of the Jacobi elliptic functions,
see \citet{CRS-1969}.


The first integrals of equations (\ref{system_non_lin_tilde}) read
\begin{equation}
\label{Manley_Rowe}
\begin{aligned}
\tilde A_2^2 - \tilde A_1^2 = \tilde A_2^2(0) - \tilde A_1^2(0), \\
\tilde A_3^2 - \tilde A_2^2 = \tilde A_3^2(0) - \tilde A_2^2(0), \\
\end{aligned}
\end{equation}
which provide the known Manley-Rowe relations and allow for a new variable 
$$
\tilde A \equiv \tilde A_1^2 - \tilde A_1^2(0) = \tilde A_2^2 - \tilde A_2^2(0) = \tilde A_3^2 - \tilde A_3^2(0),
$$
which satisfies an equation
\begin{equation}
\label{eq_tilde_A}
\partial_t \tilde A = 2 [ ( \tilde A + \tilde A_1^2(0) ) ( \tilde A + \tilde A_2^2(0) ) 
( \tilde A + \tilde A_3^2(0) ) ]^{1/2}.
\end{equation}
In the most simple case of real and positive $\tilde A_{1,2,3}$ they are expressed as
$$
\tilde A_{1,2,3} = \sqrt{ \tilde A + \tilde A^2_{1,2,3}(0)}.
$$

The solution of equation (\ref{eq_tilde_A}) reads 
\begin{equation}
\label{sol_tilde_A}
\tilde A = \frac{r_3 - r_1}{{\rm sn}^2 [(t_e-t)(r_3-r_1)^{1/2},m]} - r_3,
\end{equation}
where 
\begin{equation}
m = \frac{(r_3-r_2)^{1/2}}{(r_3-r_1)^{1/2}},
\end{equation}
\begin{equation}
\label{t_e}
t_e = (r_3-r_1)^{-1/2} {\rm sn}^{-1} [ (r_3-r_1)^{1/2}/r_3^{1/2},m]
\end{equation}
and it is assumed that each of $r_{1,2,3}$ equals to $\tilde A_{1,2,3}^2(0)$ in such an order that 
$r_3 > r_2 > r_1 > 0$. Since the solution (\ref{sol_tilde_A}) describes the case of real and negative $A_{1,2,3}$ only, 
the absolute values of these amplitudes are used hereafter $|A_{1,2,3}| = -A_{1,2,3}$. 
For brevity, the corresponding replacement $A_{1,2,3} \to - A_{1,2,3}$ is assumed everywhere below.

The solution (\ref{sol_tilde_A}) shows that
as $t\to t_e$, the amplitudes of ${\rm SDW}_r$, ${\rm IW}^\prime_r$ and ${\rm IW}^{\prime\prime}_r$ 
blow up to infinity irrespectively of their initial values, which is a manifestation of an explosive instability.
The particular curves of $A_{1,2,3}(t)$ produced for feasible $\tau=0.001$ and $f=0.01$ can be found in Figures \ref{fig_3} and \ref{fig_4}.
As the initial amplitudes take the equally large values, the time of explosion is rather short being much less than the characteristic settling 
time $\sim \tau^{-1}$. 
It can be seen that the time of explosion becomes moderately larger while the only one amplitude remains dominant at $t=0$. 
The dominant IW$_r$ makes the amplitudes to blow up far longer than that for dominant SDW$_r$, though, it occurs still within the settling time. 
This difference is expected from the dependence of $Q_1$ and $Q_{2,3}$ on $\tau$ and $f$, see equations (\ref{Q_1}) and (\ref{Q_2}), which 
show that SDW$_r$ interacts with either of IW$_r$ stronger than IW$_r^\prime$ interacts with 
IW$_r^{\prime\prime}$. As discussed below the equations (\ref{Q_1}) and (\ref{Q_2}), this is because
the amplitudes of IW$_r^\prime$ and IW$_r^{\prime\prime}$ are bounded by the value of settling velocity $\propto \tau$
according to the units chosen in this work, and in particular, in Figures \ref{fig_3} and \ref{fig_4}\footnote{See Section 
\ref{sec_lower} for the results obtained according to an alternative measuring of $\bf u$ in units of $c_s$ physically plausible 
in the particular case of modes and the collinear resonant triad.}. 
Also, the curves in Figures \ref{fig_3} and \ref{fig_4} show that the growth rate of $A_{1,2,3}$ increases as amplitudes approach $t\to t_e$. 
This suggests that the final stage of explosive instability becomes inconsistent with the underlying terminal velocity approximation. 
See the next Section, which elucidates this issue.

\subsection{Compliance with terminal velocity approximation}

The characteristic time of gas-dust dynamics due to interaction of resonant waves becomes smaller as $t \to t_e$, what can be seen in Figures 
\ref{fig_3} and \ref{fig_4}. Consequently, $\tau_*$ can be substantially larger than $\tau$, and the requirement $\tau \ll 1$ 
may become insufficient to satisfy the terminal velocity approximation, which is the case for non-interacting (linear) resonant waves, 
see Section \ref{sec_res}. Let $t_{ev}$ be defined as 
\begin{equation}
\label{t_ev_expl}
t_{ev} = \min \left \{ \frac{A_{1}}{\dot A_{1}}, \frac{A_{2}}{\dot A_{2}}, \frac{A_{3}}{\dot A_{3}}  \right \} = 
\min \left \{ \frac{ \tilde A_{1} }{ \dot {\tilde A}_{1} }, \frac{ \tilde A_{2} }{ \dot {\tilde A}_{2} }, 
\frac{ \tilde A_{3} }{ \dot {\tilde A}_{3} } \right \}
> 2\, \frac{\tilde A}{\dot {\tilde A}}.
\end{equation}

The solution (\ref{sol_tilde_A}) has a simple asymptotics close to the time of explosion, 
\begin{equation}
\label{sol_tilde_A_appr}
\tilde A \approx \frac{1}{(t_e-t)^2},
\end{equation}
as soon as
\begin{equation}
\label{cond_of_asympt}
\frac{1}{(t_e - t)^2} \gg r_3 - r_1
\end{equation}
and additionally 
\begin{equation}
\label{cond_of_asympt_add}
r_3 - r_1 \sim O(r_3).
\end{equation}
At the same time, the inverse Jacobi elliptic function in equation (\ref{t_e}) 
takes value of order of unity under the condition (\ref{cond_of_asympt_add}), what leads to an order-of-magnitude estimate
\begin{equation}
\label{relation_t_e}
t_e \sim (r_3 - r_1)^{-1/2}.
\end{equation}
Equation (\ref{relation_t_e}) implies that the restriction (\ref{cond_of_asympt}) is equivalent to 
\begin{equation}
\label{relation_t_e_1}
t_e \gg t_e-t.
\end{equation}

Thus, it follows from equation (\ref{sol_tilde_A_appr}) that $t_{ev} \gtrsim t_e - t$ under the restriction (\ref{relation_t_e_1}).
An explosive growth of resonant waves under the terminal velocity approximation requires that $t_{ev} \gg t_s$, which, in turn, implies the 
corresponding necessary condition
\begin{equation}
\label{final_cond_t_e_s}
\frac{t_s}{t_e} \ll 1. 
\end{equation}
This condition guarantees that the significant stage of an explosive growth of 
resonant triad proceeds under the terminal velocity approximation. 
On the other hand, the consideration above shows that the terminal velocity approximation is always violated sufficiently close to $t_e$, 
what occurs when $t_e-t \sim t_s$ and the corresponding growth factor of resonant waves attains the order of $t_e/t_s$.

\section{Conservation of energy}
\label{sec_energy}

The displacement for perturbations of gas-dust mixture can be introduced through the common kinematic relation 
with its centre-of-mass velocity
\begin{equation}
\label{xi_u}
\frac{d\xi}{dt} = \bf{u} + (\xi\cdot \nabla) {\bf U}.
\end{equation}

Equation (\ref{xi_u}) yields the projections
\begin{equation}
\label{xi_x}
\partial_t \xi_x = u_x,
\end{equation}
\begin{equation}
\label{xi_y}
\partial_t \xi_y = u_y - q\xi_x,
\end{equation}
\begin{equation}
\label{xi_z}
\partial_t \xi_z = u_z,
\end{equation}
which allow one to rewrite dynamical equations for linear gas-dust perturbations in terms of the 
displacement. Explicitly,
\begin{equation}
\label{eq_11}
\partial_{tt} \xi_x + \tilde\kappa^2 \xi_x = c_s^2 \partial_x (\nabla\cdot \xi),
\end{equation}
\begin{equation}
\label{eq_12}
\partial_{tt} \xi_z  = c_s^2 \partial_z (\nabla\cdot \xi) - \frac{f}{\tau} \partial_z D,
\end{equation}
\begin{equation}
\label{eq_13}
\partial_{tz} D = \tau \tilde\kappa^2 \partial_z \xi_z + (1-f) \partial_{zz} D,
\end{equation}
where $\nabla \cdot \xi \equiv \partial_x \xi_x + \partial_z \xi_z$ and it was additionally assumed that $p^\prime = c_s^2 \rho^\prime_g$ 
with $\rho_g^\prime$ being the perturbation of the gas density. 
Note that equations (\ref{eq_11}-\ref{eq_13}) are considered in the limit of incompressible dynamics, i.e.
it is assumed that $c_s\to \infty$, whereas $\rho_g^\prime \to 0$ and $\nabla \cdot \xi = 0$ 
leaving the pressure term in RHS of 
equations (\ref{eq_11}) and (\ref{eq_12}) finite. Equations (\ref{eq_12}-\ref{eq_13}) contain a new variable,
by definition,
\begin{equation}
\label{D_eq}
\partial_z D \equiv \delta.
\end{equation}

Equations (\ref{eq_11}-\ref{eq_13}) follow from the requirement that the action
\begin{equation}
\label{action}
S = \int {\cal L} (\chi^i, \partial_k \chi^i) d^3 {\bf x} dt
\end{equation}
with $d^3 {\bf x} \equiv dxdydz$ and the Lagrangian density ${\cal L}$
be stationary with respect to arbitrary variations of $\chi^i$, which is defined as 
$\chi^i \equiv \{ \xi_x, \xi_z, D \}$ in this Section.
If so, equations (\ref{eq_11}-\ref{eq_13}) are identical to the corresponding Euler-Lagrange equations
produced by the Lagrangian 
\begin{equation}
\label{Lagr}
\begin{aligned}
{\cal L} = &\frac{1}{2} \left [ (\partial_t \xi_x)^2  + (\partial_t \xi_z)^2 - \tilde \kappa^2 \xi_x^2 \right ] -  \\
&\frac{c_s^2}{2} \left [ (\partial_x \xi_x)^2 + 2 \partial_x \xi_x \partial_z \xi_z  + (\partial_z \xi_z)^2  \right ] + \\
&\frac{f}{\tau} \left [ (1-f)\, \frac{\delta^2}{2\tau \tilde\kappa^2} - 
\frac{\delta\, \partial_t D}{2\tau\tilde\kappa^2} + D\, \partial_z \xi_z \right ].
\end{aligned}
\end{equation}
Note that the thermal terms in the second square brackets in equation (\ref{Lagr}) vanish in the considered 
incompressible limit.

The symmetry of ${\cal L}$ with respect to translations in time leads to conservation of energy, 
$E \equiv \int {\cal E} d^3 {\bf x}$, where the energy density of perturbations, $\cal E$, is introduced as
$$
{\cal E} = -L + \frac{\delta {\cal L}}{\delta (\partial_t \chi_i)} \partial_t \chi_i.
$$
Equation (\ref{Lagr}) yields
\begin{equation}
\label{En}
{\cal E} = \frac{u_x^2}{2} + \frac{2 u_y^2}{\tilde\kappa^2} + \frac{u_z^2}{2} - 
\frac{f}{\tau^2 \tilde\kappa^2} (1-f) \frac{\delta^2}{2} - \frac{f}{\tau} D\, \partial_z \xi_z,
\end{equation}
which is averaged over the mode phase in order to obtain the energy density of a plane wave,
\begin{equation}
\label{En_mode}
\hat{\cal E} = \frac{\hat u_x^2}{4} + \frac{\hat u_y^2}{\tilde\kappa^2} + \frac{\hat u_z^2}{4} -  
\frac{f}{\tau^2 \tilde\kappa^2} (1-f) \frac{\hat\delta^2}{4} - \frac{f}{\tau \omega} \frac{ \hat\delta \hat u_z}{2}.
\end{equation}

To leading order in $f$, equation (\ref{En_mode}) gives the following expressions 
\begin{equation}
\label{En_i}
\begin{aligned}
&\hat{\cal E}_1 \approx - \frac{f}{\tau^2 \tilde\kappa^2} \frac{A_1^2}{4} + O(f^2),\\
&\hat{\cal E}_{2,3} \approx \frac{8\tilde\kappa^2}{k_z^2} \frac{A_{2,3}^2}{4} + O(f) \\
\end{aligned}
\end{equation}
for SDW$_r$, IW$^\prime_r$ and IW$^{\prime\prime}_r$, respectively.

Equations (\ref{En_i}) combined with equations (\ref{system_non_lin}) show that the total energy 
of the resonant triad, $\hat{\cal E}_1 + \hat{\cal E}_2 + \hat {\cal E}_3$, is conserved during
the explosive growth of waves. Therefore, explosive instability of the triad (\ref{triad}) is driven
by the conservative transfer of energy from SDW$_r$ to both IW$^\prime_r$ and IW$^{\prime\prime}_r$.

\section{Time of explosion}
\label{sec_t_e}

\begin{figure}
\begin{center}
\includegraphics[width=7.5cm,angle=0]{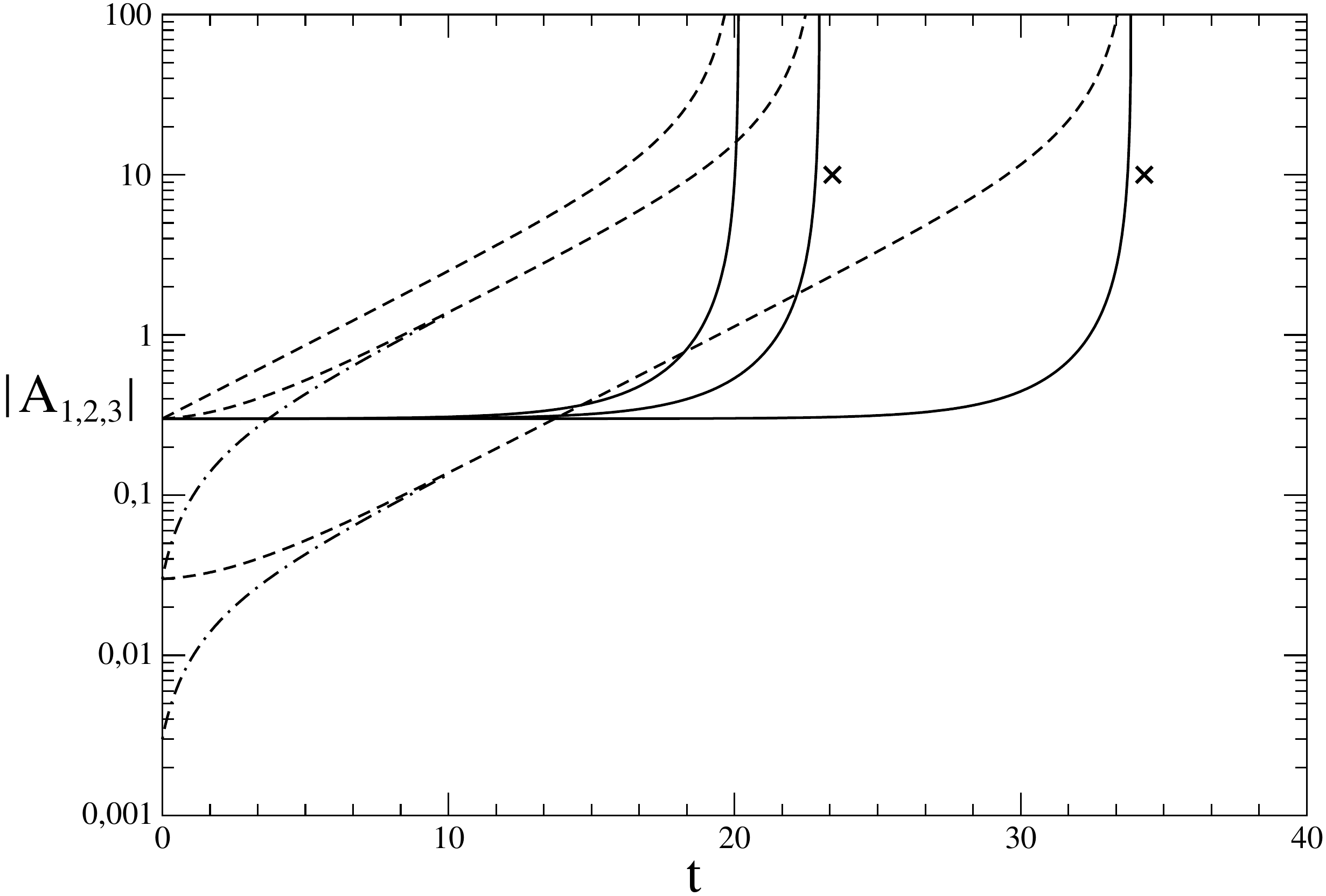}
\end{center}
\caption{Behaviour of the interacting resonant modes according to the solution (\ref{sol_tilde_A}). The amplitudes 
$A_1$, $A_2$ and $A_3$ are shown vs. time for their various initial values. Solid, dashed and dot-dashed curves stand for SDW$_r$, IW$_r^\prime$
and IW$_r^{\prime\prime}$, respectively.
Upon the increase of the time of explosion: $\{A_1(0)=0.3, A_2(0)=0.3, A_3(0)=0.3\}$, $\{A_1(0)=0.3, A_2(0)=0.3, A_3(0)=0.03\}$, 
$\{A_1(0)=0.3, A_2(0)=0.03, A_3(0)=0.003\}$. The other parameters are $\tilde \kappa = 1$, 
$k_x=k_z=\sqrt{2}$, $a=0.4$, $\tau=0.001$, $f=0.01$. The left (right) cross 
represents estimate of $t_e$ according to equation (\ref{t_e_1}) taken for $A_1(0)=0.3$ and $A_2(0)=0.3 (0.03)$.
} \label{fig_3}
\end{figure}

\vspace{0.5cm}

\begin{figure}
\begin{center}
\includegraphics[width=7.5cm,angle=0]{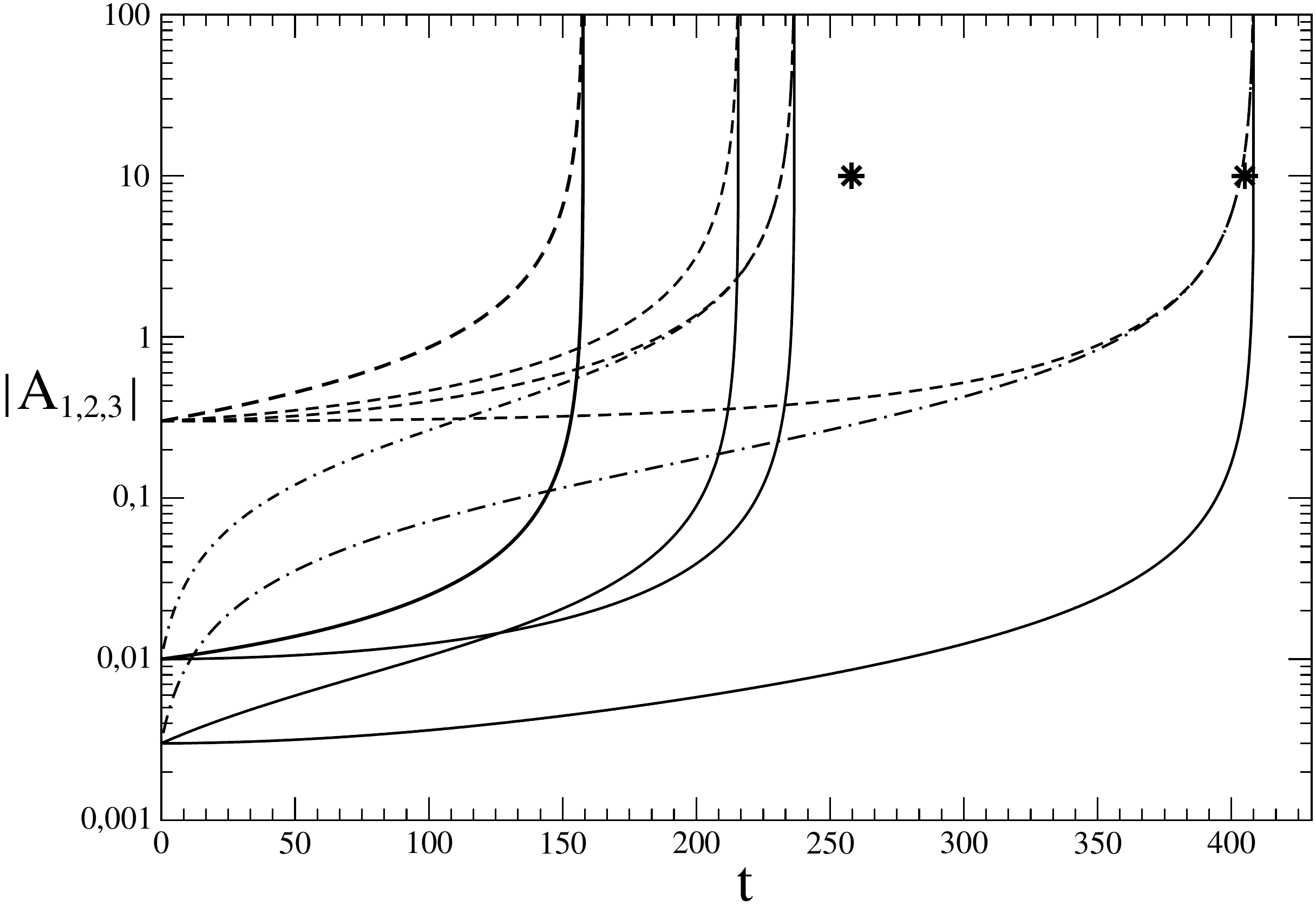}
\end{center}
\caption{Behaviour of the interacting resonant modes according to the solution (\ref{sol_tilde_A}). The amplitudes 
$A_1$, $A_2$ and $A_3$ are shown vs. time for their various initial values. Solid, dashed and dot-dashed curves stand for SDW$_r$, IW$_r^\prime$
and IW$_r^{\prime\prime}$, respectively.
Upon the increase of the time of explosion: $\{A_1(0)=0.01, A_2(0)=0.3, A_3(0)=0.3\}$, $\{A_1(0)=0.003, A_2(0)=0.3, A_3(0)=0.3\}$, 
$\{A_1(0)=0.01, A_2(0)=0.3, A_3(0)=0.01\}$, $\{A_1(0)=0.003, A_2(0)=0.3, A_3(0)=0.003\}$. The other parameters are $\tilde \kappa = 1$, 
$k_x=k_z=\sqrt{2}$, $a=0.4$, $\tau=0.001$, $f=0.01$. The left (right) star represents estimate of $t_e$ according to equation (\ref{t_e_2}) taken for $A_1(0)=0.01 (0.003)$ and $A_2(0)=0.3$.
} \label{fig_4}
\end{figure}

\begin{figure}
\begin{center}
\includegraphics[width=8cm,angle=0]{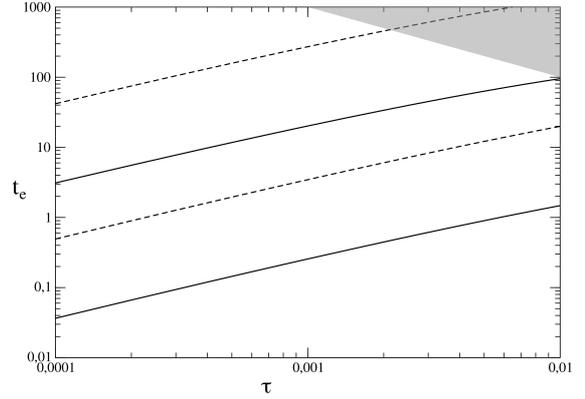}
\end{center}
\caption{The time of explosion is shown according to equation (\ref{t_e}) for the initial values of amplitudes $A_1(0)=0.3$, $A_2(0)=0.3$, $A_3(0)=0.3$
and $\tilde \kappa = 1$, $k_x=k_z=\sqrt{2}$.
For solid and dashed curves $a=0.4$ and $a=0.2$, respectively. 
In both cases, the upper and the lower curves represent, respectively, $f=0.01$ and $f=0.1$.
In the shadowed region the time of explosion becomes larger than 
the characteristic settling time of the particles. 
} \label{fig_5}
\end{figure}

The relevance of an explosive solution (\ref{sol_tilde_A}) in protoplanetary discs can be assessed 
by comparing the time of explosion, $t_e$, with characteristic timescales in a disc.

Exact values of $t_e$ in units of $\Omega_0^{-1}$ according to equation (\ref{t_e}) are shown in Figure \ref{fig_5} 
for the particular case of the amplitudes of resonant waves equal to each other. 
As expected, the time of explosion decreases for smaller particles, see discussion below the equations (\ref{Q_1}) and (\ref{Q_2}).
Also, there is a sharp inverse dependence of $t_e$ on the dust fraction.
As $f$ increases up to 0.1, it becomes shorter than the Keplerian time.
Additionally, $t_e$ decreases for IW$_r$ approaching the linear resonance between IW and SDW, $a\to 1/2$. Note that the analytical derivation 
of interaction between the resonant waves breaks as $a \to 1/2$, see text below in this Section for the corresponding estimates.

\subsection{Analytical approximations of $t_e$}

Simple estimates of $t_e$ can be made in the limiting case when the amplitudes of resonant waves strongly differ from each other at $t=0$. 
That is, there is one dominant mode, while one of the rest minor modes prevails the other one.   
In this case, $r_3 \gg r_2 \gg r_1$ and the Jacobi function standing in equation (\ref{t_e}) exhibits approximately logarithmic growth 
giving $t_e \approx r_3^{-1/2} \ln \left ( {4 r_3^{1/2}}/{r_2^{1/2}} \right )$.
Accordingly, the following general approximations of $t_e$ are obtained.

\subsubsection{Dominant SDW$_r$ and sub-dominant IW$_r^\prime$}
\label{subsec_1}

\begin{equation}
\label{t_e_1}
t_e \approx t_e^{\rm SDW} \equiv \frac{3\pi}{A_1(0)} \frac{\tau}{f^2} \frac{(2a-1)^2}{a(1-a)} \frac{k_z^2}{20 \tilde\kappa^2 + k_z^2} \, L^{\rm SDW},
\end{equation}
where
\begin{equation}
\label{L_1}
L^{\rm SDW} = \ln \left ( \frac{A_1(0)}{A_2(0)} \frac{f^{1/2}}{\tau} \frac{k_z}{\tilde\kappa^2}  \right ) 
\end{equation}
provided that 
$$
A_2(0) \gg A_3(0)
$$ 
and 
$$
\frac{A_1(0)}{A_2(0)} \gg \frac{4\tau}{f^{1/2}} \frac{\tilde \kappa^2}{k_z}.
$$

Note that as far as $\tau$ is sufficiently small, this regime can be valid also for comparable $A_1(0) \sim A_2(0)$, see
the accordance of equation (\ref{t_e_1}) with an exact analytical solution in Figure \ref{fig_3}. 
The case of sub-dominant IW$_r^{\prime\prime}$ is considered similarly with the replacements $A_{2,3}(0) \to A_{3,2}(0)$.


\subsubsection{Dominant IW$^{\prime}_r$ and subdominant SDW$_r$}
\label{subsec_2}

\begin{equation}
\label{t_e_2}
t_e \approx t_{e\, 1}^{\rm IW} \equiv 
\frac{3\pi}{4 A_{2}(0)} \frac{1}{f^{3/2} \tilde\kappa^2} \frac{(2a-1)^2}{a(1-a)} \frac{k_z^3}{20 \tilde\kappa^2 + k_z^2} \, L_1^{\rm IW},
\end{equation}
where 
\begin{equation}
\label{L_2}
L_1^{\rm IW} = \ln \left ( 16 \frac{A_2(0)}{A_1(0)} \frac{\tau}{f^{1/2}} \frac{\tilde\kappa^2}{k_z}  \right ) 
\end{equation}
provided that 
$$
\frac{A_1(0)}{A_3(0)} \gg \frac{4\tau}{f^{1/2}} \frac{\tilde \kappa^2}{k_z}.
$$
and
$$
\frac{A_2(0)}{A_1(0)} \gg \frac{f^{1/2}}{4\tau} \frac{k_z}{\tilde \kappa^2}.
$$

Note that as far as $\tau$ is sufficiently small, this regime can be valid also for comparable $A_1(0) \sim A_3(0)$, 
while $A_2(0)$ should significantly exceed $A_1(0)$, see the accordance of equation (\ref{t_e_2}) with an exact analytical 
solution in Figure \ref{fig_4}.

\subsubsection{Dominant IW$^{\prime}_r$ and sub-dominant IW$^{\prime\prime}_r$}
\label{subsec_3}

\begin{equation}
\label{t_e_2_2}
t_e \approx t_{e\, 2}^{\rm IW} \equiv 
\frac{3\pi}{4 A_{2}(0)} \frac{1}{f^{3/2} \tilde\kappa^2} \frac{(2a-1)^2}{a(1-a)} \frac{k_z^3}{20 \tilde\kappa^2 + k_z^2}\, L_2^{\rm IW},
\end{equation}
where 
\begin{equation}
\label{L_3}
L_2^{\rm IW} = \ln \left ( 4 \frac{A_2(0)}{A_3(0)} \right ) 
\end{equation}
provided that 
$$
\frac{A_3(0)}{A_1(0)} \gg \frac{f^{1/2}}{4\tau} \frac{k_z}{\tilde \kappa^2}.
$$
and
$$
\frac{A_2(0)}{A_3(0)} \gg 1.
$$

Consideration of dominant IW$^{\prime\prime}_r$ is identical to Sections \ref{subsec_2} and \ref{subsec_3}  
with the replacements $A_{2,3}(0) \to A_{3,2}(0)$.

\subsection{Lower estimates of $t_e$}
\label{sec_lower}

As expected, $t_e$ becomes shorter as the inverse initial amplitude of the dominant mode. 
At the same time, it increases in gas-dust mixture with smaller dust fraction, 
however, becoming shorter for smaller particles\footnote{For dominant IW$_r$ the latter is true provided that 
its amplitude is measured in units of $c_s \equiv g_z/\Omega_0$ rather than $g_z t_s$, i.e. $A_{2,3} \to A_{2,3}/\tau$ 
in equation (\ref{t_e_2}) and (\ref{t_e_2_2}). \label{f_n}}, see the discussion below the equations (\ref{Q_1}-\ref{Q_2}).

The time of explosion can be additionally decreased for the triads containing  
inertial waves located closer to the band of DSI, so for $a\to 1/2$. Alternatively, 
time of explosion decreases for almost radially propagating resonant modes, $k_z \ll 1$.
The corresponding lower estimate of $t_e$ can be obtained using the marginal condition of the validity of 
the analytical approximation employed in this work.
Namely, the analytical form of waves involved in the resonant triad is valid under the restrictions 
(\ref{iw_appr}-\ref{sdw_appr}) and (\ref{iw_not_res}-\ref{sdw_not_res}). 
It can be checked that taken at the resonant wavenumbers they are satisfied together provided 
that the overall condition 
\begin{equation}
\label{overall_restr}
\frac{4f\tilde\kappa^2}{(1-2a)^2} \frac{k_x^2}{k_z^2} \ll 1
\end{equation}
is true.
An additional restriction comes from the condition that 
$\Delta k / k \ll  1$.  Equation (\ref{three_wave_cond}) yields
\begin{equation}
\label{delta_k_restr}
\frac{8f \tilde\kappa^2}{1-2a} \left ( \frac{1}{k_x^2} + \frac{5}{6} \frac{1}{k_z^2} \right )  \ll 1.
\end{equation}

The restrictions (\ref{overall_restr}) and (\ref{delta_k_restr}) put the lower limit on the value of $|2a-1| k_z$, 
or alternatively, on the values of $|2a-1|$ and $k_z$ separately, which 
enter the numerators of equations (\ref{t_e_1}), (\ref{t_e_2}) and (\ref{t_e_2_2}).
They provide the lower estimates of $t_e^{\rm SDW}$ and $t_{e\,1,2}^{\rm IW}$. It is convenient to formulate these
estimates separately for different cases defined by the ratio between $k_x$ and $k_z$.

Note that logarithmic factors entering equations (\ref{t_e_1}), (\ref{t_e_2}) and (\ref{t_e_2_2}) are omitted in the following estimates,
thus,   $t_{e\,1}^{\rm IW} \approx t_{e\,2}^{\rm IW} \equiv t_{e}^{\rm IW}$.

Hereafter it is assumed that $A_{2,3}(0)$ is measured in units of $c_s \equiv g_z/\Omega_0$, 
see discussion below the equations (\ref{Q_1}-\ref{Q_2}) and the footnote \ref{f_n}.

\subsubsection{Almost radially propagating modes $k_z \ll k_x$}

\begin{equation}
\label{t_e_11}
\begin{aligned}
& t_{e}^{\rm SDW} \gtrsim \frac{48\pi}{5 A_1(0)} \frac{\tilde\kappa^2 \tau}{f}, \\
& t_{e}^{\rm IW} \gtrsim \frac{24\pi}{A_{2,3}(0)} \tilde\kappa^2 \tau.
\end{aligned}
\end{equation}
In equation (\ref{t_e_11}) it is assumed that combinations of $a$ entering denominators of 
(\ref{t_e_1}) and (\ref{t_e_2}) take approximately their largest values at $0<a<1/2$.   

There should be a caution about the lower estimate of $t_e^{\rm IW}$ 
from equation (\ref{t_e_11}), which formally provides the existence of an explosive instability for $f\to 0$.
The inspection of the coupling coefficients shows that this issue originates from the divergence 
of $Q_1$ as soon as $f\to 0$ and $k_z \sim O(f^{1/2})$, which is marginally allowed 
by the non-resonant linear solution for IW$_r$ in the case of almost radially propagating waves.
As was discussed in Section \ref{sec_coupling}, the interaction between IW$_r^\prime$ and IW$_r^{\prime\prime}$
is produced mainly by the velocity perturbation of one IW$_r$ and the aerodynamically induced perturbation of 
the dust density of the other IW$_r$. However, equation (\ref{lin_eq_4}) indicates that the latter diverges
in the limit $k_z\to 0$. Such a singularity should be removed with the account of the dust diffusion.
The further study of an explosive instability should check that explosion time of the resonant triad consisting of the radially 
propagating waves tends to infinity as $f\to 0$ in the system with non-zero dust diffusion.

\subsubsection{Almost vertically propagating modes $k_x \ll k_z$}

\begin{equation}
\label{t_e_21}
\begin{aligned}
& t_e^{\rm SDW} \gtrsim \frac{8\pi}{A_1(0)} \frac{\tilde\kappa^{4/3} \tau}{f^{2/3}}, \\
& t_e^{\rm IW} \gtrsim \frac{4\pi}{A_{2,3}(0)} \frac{\tilde\kappa^{1/3} \tau}{f^{1/6}}
\end{aligned}
\end{equation}
From the derivation of equation (\ref{t_e_21}) it follows that $a$ can be approximately set
to $1/2$ in this case.

\subsubsection{Modes with $k_x \approx k_z$}

\begin{equation}
\label{t_e_31}
\begin{aligned}
& t_e^{\rm SDW} \gtrsim \frac{48\pi}{11 A_1(0)} \frac{\tilde\kappa^2 \tau}{f}, \\
& t_e^{\rm IW} \gtrsim \frac{48\sqrt{2}\pi}{44A_{2,3}(0)} \frac{\tilde\kappa \tau}{f^{1/2}}.
\end{aligned}
\end{equation}
From the derivation of equation (\ref{t_e_31}) it follows that $a$ can be approximately set
to $1/2$ in this case.

It can be seen that for reasonable values of $f$ 
the estimates (\ref{t_e_21}) give the least lower limits on the time of explosion in 
gas-dust mixture with small dust fraction. Equation (\ref{t_e_21}) is used below in the next Section.


\subsection{Keplerian disc}
\label{sec_Keplerian}

The lower bound of the time of explosion over all limiting cases given above can be 
estimated in a Keplerian disc, where $\tilde \kappa = 1$, as the following 
\begin{equation}
\label{fin_t_e}
\begin{aligned}
& t_e^{\rm SDW}  \gtrsim \tilde t_e^{\rm SDW} \equiv 20 \frac{\tau}  {A_{1}(0) f^{2/3} }, \\
& t_e^{\rm IW}  \gtrsim \tilde t_e^{\rm IW} \equiv 10 \frac{\tau} {A_{2,3} (0) f^{1/6} },
\end{aligned}
\end{equation}
which is measured in units of the Keplerian time for initially dominant SDW$_r$ and IW$_r$, respectively.
This choice corresponds to waves propagating almost vertically.

\subsubsection{On excitation of dominant IW$_r$}
\label{sec_excite_EI}

Equations (\ref{fin_t_e}) indicate that the case of dominant IW$_r$ looks 
preferable to the case of dominant SDW$_r$ with respect to transition to an explosive instability, 
because of quite a weak dependence of the time of explosion on the dust fraction.  
Therefore, the question arises about an excitation of IW$_r$ with the amplitude sufficient to 
trigger explosive instability.

One possibility is a preliminary linear growth of IW$_r$ due to DSI. 
However, the resonant triad considered analytically in this work contains IW$_r$ located 
far from the linear resonance between IW and SDW, which gives rise to the leading order DSI, 
see the corresponding condition (\ref{iw_not_res}). 
Therefore, the leading order DSI cannot be responsible 
for the production of such a finite-amplitude IW$_r$. 
Nevertheless, exact solution of the general dispersion equation in \cite{squire_2018} shows that  
the particular curves of DSI growth rate have broad wings 
of the growth rate outside of the main band of DSI associated with the linear resonant coupling 
between SDW and IW. The following remains to be checked, however,
such a 'residual' linear instability may occur due to some additional
mechanism responsible for slow growth of the upcoupled IW and SDW in the non-resonant range of wavenumbers 
satisfying the condition (\ref{iw_not_res}). If so, IW at these wavenumbers may become subject for further 
explosive growth due to the non-linear resonant interaction with the corresponding seeded SDW and other IW 
having much smaller amplitudes. On the other hand, the condition (\ref{iw_not_res}) can be relaxed 
in the sequel studies of an explosive instability, in which case the coupling coefficients for 
three-wave resonance can be obtained numerically. Provided that explosive instability keeps its strength for 
IW also from the band of the leading order DSI considered by \citetalias{zhuravlev-2019}, 
it may be exactly the non-linear stage of DSI.  

Besides, IW are known to be excited in turbulent rotating fluids. 
This process has long been observed and simulated in the laboratory tanks, 
see e.g. \citet{hopfinger-1982}, \citet{godeferd-1999}, \citet{bewley-2007}, \citet{lamriben-2011}.
Some kind of turbulence pre-existing in the gas component including those generated by the shear of Keplerian 
motion may also be responsible for generation of finite amplitude IW subject to explosive instability.
The energy spectrum of such waves in a disc must be a special issue. However, Kolmogorov cascade seems to be
unsuitable to trigger an explosive instability. Indeed, the usual assumption that turbulence is characterised 
by the largest velocity fluctuations $V_t \sim \alpha^{1/2} c_s $, at the outer scale $L_t \sim \alpha^{1/2}h$, 
where $h$ and $\alpha$ are the disc scaleheight and the Shakura-Sunyaev parameter, 
leads to the turbulent velocity fluctuations $v_t \sim \alpha^{1/3} \tau^{1/3} c_s$ 
and its correlation time $t_{corr} \sim \alpha^{-1/3} \tau^{2/3} \Omega_0^{-1}$ evaluated at the scale of DSI, which 
is $k_{\rm DSI} \sim 1/(\tau h)$\footnote{It is assumed for simplicity that scale of dominant IW$_r$ is of order of $k_{\rm DSI}$.}
in the dimensional form. If the amplitude entering $\tilde t_e^{\rm IW}$ from equation (\ref{fin_t_e}) is supposed to be identified with $v_t$, 
the time of explosion of the corresponding explosive instability is found to be larger than $t_{corr}$. The latter implies that 
IW should disappear due to interaction with other modes of turbulent cascade before it could be amplified by some resonant SDW. 
Note that such a turbulent excitation of explosive instability would be possible at the sufficiently small scales 
corresponding to $\tau \lesssim \alpha^{1/2}$ as seen by comparing $L_t^{-1}$ with $k_{\rm DSI}$. At large scales, $\tau \gtrsim \alpha^{1/2}$, 
turbulence has a damping effect. In this regime, the corresponding time of linear damping, $t_\nu$, must be larger than $t_{corr}$. The 
next Section is devoted to evaluation of the lower $\tau = \tau_\nu$ corresponding to damping of explosive instability 
as far as $t_\nu \gtrsim t_{corr}$.

\subsubsection{Viscous threshold for explosive instability}

The threshold $\tau \equiv \tau_\nu$ corresponding to damping of an explosive instability by 
dissipative processes in a disc can be estimated using the corresponding an order-of-magnitude condition
known from the theory of resonant interaction between waves with linear damping, see e.g. 
\citet{wilhelmsson-1970_1} and \citet{wilhelmsson-1970_2}. As soon as $t_e \gtrsim t_\nu$, where $t_\nu$
is the characteristic time of linear damping, an explosive growth of waves does not exist anymore.
The time of linear damping on the scale of resonant triad reads
\begin{equation}
\label{t_nu}
t_\nu \simeq \frac{\tau^2}{\alpha},
\end{equation}
where it is assumed that $\alpha$ characterises the disc effective viscosity via the 
common relation for kinematic viscosity $\nu = \alpha  \Omega_0 h^2$.
Note that equation (\ref{t_nu}) is obtained according to the assumption that $g_z \simeq \Omega_0^2 h$, 
so takes its maximum value in a disc, which provides the largest $t_\nu$ for the given $\tau$.
Accordingly, the smallest particles and the corresponding lower bound of the resonant length-scales 
subject to explosive instability in a viscous disc are introduced by
\begin{equation}
\label{tau_nu}
\tau_\nu \simeq \frac{10 \alpha}{A_{2,3}(0) f^{1/6}},
\end{equation}
which is obtained equating $\tilde t_e^{\rm IW}$ from equation (\ref{fin_t_e}) with $t_\nu$. 
The corresponding lower bound for $\tilde t_e^{\rm IW}$ reads
\begin{equation}
\label{t_e_nu}
\tilde t_{e\,\nu}^{\rm IW} \simeq \frac{100 \alpha}{A^2_{2,3}(0) f^{1/3}}.
\end{equation}


\subsubsection{Settling threshold for explosive instability}

Conversely, the biggest particles and the corresponding upper bound of the resonant length-scales 
subject to explosive instability are specified by the restriction that the time of explosion 
cannot be larger than settling time of the particles, $t_{stl} \simeq \tau^{-1}$. Equating $t_{stl}$ with 
$\tilde t_e^{\rm IW}$ from equation (\ref{fin_t_e}) one obtains the corresponding largest
$\tau$ as the following
\begin{equation}
\label{tau_stl}
\tau_{stl} \simeq \frac{A_{2,3}^{1/2}(0) f^{1/12}}{10^{1/2}}.
\end{equation}
The corresponding upper bound for $\tilde t_e^{\rm IW}$ reads
\begin{equation}
\label{t_stl}
\tilde t_{stl} \simeq \frac{10^{1/2}}{A_{2,3}^{1/2}(0) f^{1/12}}.
\end{equation}

\subsubsection{Threshold for an amplitude exciting explosive instability}

Generally, $\tilde t_e^{\rm IW}$ and the corresponding $\tau$ cover the range
\begin{equation}
\label{ranges_tau_t_e}
\begin{aligned}
&\tau_{stl} \gtrsim \tau \gtrsim \tau_\nu, \\
& \tilde t_{stl} \gtrsim \tilde t_e^{\rm IW} \gtrsim \tilde t_{e \,\nu}^{\rm IW}. 
\end{aligned}
\end{equation}

In a weakly viscous disc with small $\alpha$ as well as for $A_{2,3}(0)$ close to unity, 
$\tau$ covers a range from the value much smaller than one up to the value slightly less than one. 
As the initial amplitude decreases, the allowed area for $\tau$ subject to an explosive instability reduces. 
The lower threshold value of $A_{2,3}(0)$ subject to an explosive instability
is estimated from the condition $\tau_{stl} \simeq \tau_\nu$:
\begin{equation}
\label{A_cr}
A_{2,3}(0) \gtrsim \frac{10 \alpha^{2/3}}{f^{1/6}}.
\end{equation}
Below this value the particular case of explosive instability 
considered in this work cannot exist in a viscous gas-dust medium.
The corresponding marginal values of $\tau$ and $\tilde t_{e}^{\rm IW}$ are, respectively, the following
\begin{equation}
\label{tau_t_e_cr}
\begin{aligned}
& \tau \simeq \alpha^{1/3},
& \tilde t_{e}^{\rm IW} \simeq \alpha^{-1/3}.
\end{aligned}
\end{equation}

Note that the formal restriction of the weakly non-linear theory, $A_{2,3}(0) \lesssim 1$, 
provides the upper limit of the viscosity parameter, 
\begin{equation}
\label{alpha_max}
\alpha \lesssim \frac{f^{1/4}}{10^{3/2}} \simeq 0.01,
\end{equation}
obtained from equation (\ref{A_cr}) for $f \simeq 0.01$. 
This restriction also follows from the condition that $\tau_{stl} \gtrsim \tau_\nu$ 
or $\tilde t_{stl} \gtrsim \tilde t_{e\, \nu}^{\rm IW}$ taken with $A_{2,3}(0) \simeq 1$.
Viscous discs where the restriction (\ref{alpha_max}) is violated are stable with respect to an explosive instability.


\section{Conclusions}
\label{sec_final}

This work is focused on possibility of weakly non-linear instability of gas-dust 
mixture with the dust settling through the horizontally rotating gas under the vertical hydrostatic equilibrium. 
It is revealed that such a flow is subject to explosive instability provided that the dust back reaction 
on gas is taken into account. The physics of instability is considered through the particular example of 
three-wave resonance among axisymmetric gas-dust waves, which are the counterparts of one SDW and two IW 
modified by the small amount of dust in a mixture. 
The fundamental reason that causes an explosive instability is the energy of SDW, 
which becomes negative provided that the dust settling is sufficiently fast, 
see \citetalias{zhuravlev-2019}. This enables the unbounded growth of resonant waves, while the energy 
is transferred from SDW to two IW. At the same time, it is shown that interaction between the waves 
conserves the total energy of the resonant triad. 

The main application of the considered model is a small patch of protoplanetary disc above the disc
midplane. However, it can also be applied to other situations with the rotational profile different 
from the Keplerian one. For example, the considered model may be applied to local environments of 
the dust-laden envelopes of the rotating giant planets forming through the pebble accretion, 
see e.g., \citet{lambrechts-2017}. 
Explosive instability of gas-dust mixture can be one more physical effect that accompanies complicated process 
of accretion of small solids inside the envelope, for the recent account 
see e.g. \citet{popovas-2018}, \citet{nordlund-2020} and references therein.
Indeed, the only parameter here, which describes the deviation from the rigid rotation 
is the dimensionless epicyclic frequency changing from two to one while replacing 
the rigid rotation by the Keplerian rotation. 
The coupling coefficients obtained for the resonant triad keep their negative signs, or equivalently, 
an explosive type irrespectively of the rotational profile, 
whereas the analytical estimates exposed in Section \ref{sec_t_e} show that the time of explosion 
weakly depends on epicyclic frequency. Along with DSI, this makes explosive instability a generic process in 
rotating dusty astrophysical flows. 

This work deals with particularly simple variant of resonant triad, which consists of collinear waves as $f\to 0$ and
allows for fully analytical treatment. However, Section \ref{sec_res} also introduces the general resonant triad spanning 
a wide range of wavenumbers, $(0,2\tilde \kappa)$. Derivation of the non-linear coupling between the corresponding resonant waves 
is relegated to the future work. However, it should produce explosive instability widespread in phase space covering the linearly 
stable wavenumbers, where DSI is absent. 

It was found that the conservative three-wave resonant interaction tends to infinity as $\tau \to 0$.
This is explained by the shift of the resonant scales to infinitely small lengths, which makes the 
characteristic 'non-linear frequency' $\sim A / l_{res}$\footnote{here $A$ and $l_{res}$ are, respectively, the 
velocity amplitude of resonant wave and resonant lengthscale.} entering the non-linear terms
of equations (\ref{eq_1}-\ref{eq_2}) diverge. In this situation, the lower spatial scale of three-wave
resonance should be defined by viscous damping. The corresponding estimates lead to the overall conclusion
that explosive instability can operate in discs with the usual dimensionless viscosity less than 0.01. 
The plausible scenario for transition of gas-dust mixture to explosive instability is briefly discussed in Section \ref{sec_excite_EI}.


Subsequent studies of explosive instability in protoplanetary discs should be expanded to other resonances
of an explosive type, which exist in gas-dust mixture with dust settling to the disc midplane.
An important issue is to understand the possible connection between DSI and explosive instability, in particular, 
whether explosive instability can serve directly as the non-linear stage of DSI. The latter would imply 
that the dust overdensities can reach values at least comparable to the background value of the dust density.
On that way, one of the necessary steps would be semi-analytical study of the particular three-wave resonance
proposed in this work with IW$_r$ located close to the linear resonance with SDW, therefore, 
inside the band of the leading order DSI. 
Also, saturation of explosive instability should be examined employing the corresponding numerical simulations. 
The more challenging task is to study the saturation of explosive instability analytically employing the weakly non-linear theory 
in the third order over the amplitudes of SDW and IW. The corresponding resonant tetrads may define the fate 
of dust clumps as they attain sufficiently large amplitudes. 

On the other side, the simple background solution used in this work should be generalised onto the settling 
of particles combined with their radial drift. 
The dispersion equation for SDW accounting for the dust radial drift 
may substantially affect the resonant triad as well as the interactions between the resonant waves. 
The settling of particles combined with their radial drift is the case considered by \citet{squire_2018} 
as they found DSI. Later on, \citetalias{zhuravlev-2019} suggested that the corresponding small scale
asymptotics of DSI exhibiting an unbounded growth rate is produced by the triple linear coupling of one
negative energy SDW with two positive energy IW. 
\citet{krapp-youdin-2020} studied dynamics of the non-linear gas-dust perturbations trying to determine the saturation level 
of the dust clumping after the linear growth of dust density perturbations caused by this branch of DSI.
At the same time, the authors report that the large scale asymptotics of DSI determined solely by the settling 
of dust requires substantially higher numerical resolution, see their Figure C1 and also the left panel in their Figure 5. The latter 
suggests that simulations do not reproduce DSI for $k_x \lesssim k_z$ and $k \sim k_{\rm DSI}$
even though 
the DSI growth rate must be of order of $\Omega_0$. Similarly, at least a longer wavelength mode of the resonant triad 
considered here may be numerically inhibited in the results of \citet{krapp-youdin-2020}.
Additionally, the start of simulations from white noise of velocity should involve IW$_r$ in interactions with multiple high-amplitude 
small-scale IW, which may suppress explosive instability similarly to the action of turbulent damping. 
Detailed analysis of the power spectrum of evolving perturbations is required in order to resolve these issues.  
Nevertheless, note that \citet{krapp-youdin-2020} obtained the dust overdensities much larger than the background value of the dust density reaching 
fully non-linear regime of the dust clumping in the majority of runs. It is possible that explosive instability operates 
at early stage of simulations. Whether its contribution to dust clumping or/and the transition to turbulence 
revealed by \citet{krapp-youdin-2020} is substantial should be addressed in the future work. 


At last, the particular analytical solution obtained in this work can serve a good test for the numerical schemes
employed to simulate the non-linear dynamics of gas-dust mixtures.

\section*{Data availability}
No new data were generated or analysed in support of this research.

\section*{Acknowledgments}
The author thanks Jonathan Squire for his careful review of the manuscript, which enabled significant improvement of its clarity.
The work was supported by the Foundation for the Advancement of Theoretical
Physics and Mathematics ``BASIS'' and  the Program of development of Lomonosov Moscow State University.
The analysis of explosive solutions was supported by 
the Ministry of Science and Higher Education of the Russian Federation grant 075-15-2020-780 (N13.1902.21.0039).

\bibliography{bibliography}

\appendix

\onecolumn

\section{Description of main variables}

\label{app_symb}

\begin{tabular}{ll}
Symbol     & Meaning    \\ 
\hline
$\rho_g$  &  gas volume density  \\

$\rho_p$  &  dust volume density \\

$\rho$ & total density of gas-dust mixture \\

$f$ & dust fraction \\

$p$  & gas pressure \\

${\bf U}_g$ & velocity of gas \\

${\bf U}_p$ & velocity of dust \\

${\bf U}$ & centre-of-mass velocity of gas-dust mixture \\ 

${\bf V}$ & relative velocity of gas-dust mixture \\

${\bf u}$ & the Eulerian perturbation of ${\bf U}$ \\

$\rho_p^\prime$ & the Eulerian perturbation of $\rho_p$ \\

$p^\prime$ & the Eulerian perturbation of $p$ \\

$\delta$ & the relative perturbation of $\rho_p$ \\

$W$ & perturbation of enthalpy of gas-dust mixture \\

$\Omega_0$ & local angular velocity of disc \\

$g_z$ & vertical component of the stellar gravity \\

$q$ & local disc shear rate \\

$\tilde \kappa$ & epicyclic frequency in units of $\Omega$ \\ 

$t_s$ & particle stopping time \\

$\tau$ & the Stokes number \\

$k_x$ & radial wavenumber of mode \\

$k_z$ & vertical wavenumber of mode \\

$k$ & absolute value of wavenumber of mode \\

$\omega$ & the mode frequency \\

$\omega_i$ & frequency of inertial wave in the limit $f\to 0$ \\

$\omega_p$ & frequency of the streaming dust wave in the limit $f\to 0$ \\

$\epsilon$ & coupling term of the linear dispersion equation \\

$\Delta_p$ &  non-resonant correction to $\omega_p$ due to the dust back reaction on gas \\ 

$\Delta_i^\pm$ &  non-resonant correction to $\omega_i$ due to the dust back reaction on gas \\

$\hat u_{x,y,z}$ & Fourier harmonics of $u_{x,y,z}$ \\

$\hat\delta$     & Fourier harmonic of $\delta$ \\

SDW$_r$ & resonant streaming dust wave \\

IW$_r^\prime$ & the first resonant inertial wave \\

IW$_r^{\prime\prime}$ & the second resonant inertial wave \\

${\bf k}_{\rm DSI}$ & wavenumber of the mode crossing, which gives rise to DSI \\

${\bf k}^\bullet$ & wavenumber of SDW$_r$ \\

${\bf k}^\prime$ & wavenumber of IW$_r^\prime$ \\

${\bf k}^{\prime\prime}$ & wavenumber of IW$_r^{\prime\prime}$ \\

$a$ & free parameter setting the resonant triad \\

$A_1$  & amplitude of the relative dust density perturbation in SDW$_r$ \\

$A_2$  & amplitude of the radial projection of ${\bf u}$ in IW$_r^\prime$ \\

$A_3$  & amplitude of the radial projection of ${\bf u}$ in IW$_r^{\prime\prime}$ \\

$Q_1$ & coefficient for non-linear coupling between IW$_r^\prime$ and IW$_r^{\prime\prime}$ \\

$Q_2$ & coefficient for non-linear coupling between SDW$_r$ and IW$_r^{\prime\prime}$ \\

$Q_3$ & coefficient for non-linear coupling between SDW$_r$ and IW$_r^{\prime}$ \\

$t_e$ & explosion time needed by $A_{1,2,3}$ to blow up to infinity \\

$t_e^{\rm SDW}$ & the value of $t_e$ for the dominant SDW$_r$ \\

$t_e^{\rm IW}$ & the value of $t_e$ for the dominant IW$_r$ \\

$\tilde t_e^{\rm SDW}$ & the lower estimate of $t_e^{\rm SDW}$ \\

$\tilde t_e^{\rm IW}$  & the lower estimate of $t_e^{\rm IW}$ \\

$c_s$ & local speed of sound \\

$h$ & the disc scaleheight \\

$\alpha$ & the Shakura-Sunyaev viscosity parameter \\

$\tilde t_{e\, \nu}^{\rm IW}$ & the lower value of $\tilde t_e^{\rm IW}$ estimated in a viscous disc \\

$\tau_\nu$ & the lower value of $\tau$ subject to explosive instability in a viscous disc \\

$\tilde t_{stl}^{\rm IW}$ & the upper value of $\tilde t_e^{\rm IW}$ comparable to the settling time \\

$\tau_{stl}$ & the upper value of $\tau$ subject to explosive instability quenched by fast  dust settling \\

\hline
\end{tabular}

Note that the superscript $\bullet$ is omitted after $k_x$ and $k_z$ throughout all Sections of the Appendix.

\section{Set of eigen-frequencies at resonant wavenumbers}
\label{freqs}

I) At the wavenumber of SDW$_r$.

i) SDW

\begin{equation}
\label{app_SDW_sdw_freq}
\omega = -k_z \left ( 1 - f - f \frac{k_x^2}{3k_z^2} \right ).
\end{equation}

ii) IW$^-$

\begin{equation}
\label{app_SDW_iwm_freq}
\omega = - \frac{k_z}{2} \left ( 1 + f \frac{k_x^2}{k_z^2} \right ). 
\end{equation}

iii) IW$^+$

\begin{equation}
\label{app_SWD_iwp_freq}
\omega = \frac{k_z}{2} \left ( 1 + \frac{f}{3} \frac{k_x^2}{k_z^2} \right ). 
\end{equation}

II) At the wavenumber of IW$^\prime_r$.

i) SDW

\begin{equation}
\label{app_IW1_sdw_freq}
\omega = - (a k_z + \Delta k_z) + f a k_z \left ( 1 + \frac{k_x^2}{k_z^2 (4a^2-1)} \right ).
\end{equation}

ii) IW$^-$

\begin{equation}
\label{app_IW1_iwm_freq}
\omega = - \frac{k_z}{2} - f\frac{k_x}{2a-1} \left [ (1-a)\left ( \frac{k_z}{k_x} + \frac{5}{6}\frac{k_x}{k_z} \right )
+ \frac{k_x}{k_z} \frac{a}{2}  \right ]. 
\end{equation}

iii) IW$^+$

\begin{equation}
\label{app_IW1_iwp_freq}
\omega = \frac{k_z}{2} + f k_x \left [ \frac{1-a}{2a-1} \left ( \frac{k_z}{k_x} + \frac{5}{6}\frac{k_x}{k_z} \right )
+ \frac{k_x}{k_z} \frac{a}{2(2a+1)}  \right ]. 
\end{equation}

At the wavenumber of IW$^{\prime\prime}_r$ the frequencies are obtained by the replacement 
$a \to 1-a$ and $\Delta k_z \to - \Delta k_z$ in equations (\ref{app_IW1_sdw_freq}-\ref{app_IW1_iwp_freq}).

\section{Set of eigen-vectors at resonant wavenumbers}
\label{basis}

%
%
%

I) At the wavenumber of SDW$_r$.

i) SDW

\begin{equation}
\left \{
\begin{aligned}
\label{app_SDW_sdw}
- {\rm i} & \frac{f}{\tau}\frac{k_x}{3 \tilde \kappa^2}, \\
& \frac{f}{\tau} \frac{k_x}{6 k_z}, \\
{\rm i} & \frac{f}{\tau} \frac{k_x^2}{3 k_z \tilde \kappa^2}, \\
& \quad 1. \quad
\end{aligned}
\right \}
\end{equation}

ii) IW$^-$

\begin{equation}
\left \{
\begin{aligned}
\label{app_SDW_iwm}
1, \quad\quad\quad \\
\frac{{\rm i \tilde \kappa^2}}{k_z} \left ( 1 - f \frac{k_x^2}{k_z^2}
\right ), \\
-\frac{k_x}{k_z}
, \quad \quad \\
\frac{4{\rm i} \tau  \tilde \kappa^2 k_x}{k_z^2} (1+2 f).
\end{aligned}
\right \}
\end{equation}

iii) IW$^+$

\begin{equation}
\left \{
\begin{aligned}
\label{app_SDW_iwp}
1, \quad\quad\quad\quad\quad\quad \\
-\frac{{\rm i  \tilde \kappa^2}}{k_z} \left ( 1 - f \frac{k_x^2}{3 k_z^2} 
\right ),\quad\quad \\
-\frac{k_x}{k_z} 
,  \quad\quad\quad\quad\quad \\
- \frac{4{\rm i} \tau \tilde \kappa^2\, k_x}{3 k_z^2} \left ( 1 - \frac{4}{9} f \frac{k_x^2}{k_z^2} + \frac{2}{3}f  
\right ).
\end{aligned}
\right \}
\end{equation}

II) At the wavenumber of IW$^\prime_r$.

i) SDW

\begin{equation}
\left \{
\begin{aligned}
\label{app_IW1_sdw}
- {\rm i} & \frac{f}{\tau} \frac{k_x}{\tilde\kappa^2}\frac {a}{4a^2-1},\quad\quad \\
&\frac{f}{\tau} \frac{k_x}{2 (4a^2-1) k_z},\quad \\
{\rm i} & \frac{f}{\tau} \frac{k_x^2}{k_z\tilde\kappa^2} \frac{a}{4a^2-1}, \\
& \quad\quad\quad 1.
\end{aligned}
\right \}
\end{equation}

ii) IW$^-$

\begin{equation}
\left \{
\begin{aligned}
\label{app_IW1_iwm}
1,\hspace{7cm} \\
\\
\frac{{\rm i\tilde\kappa^2}}{k_z} \left \{ 1 - \frac{2f k_x}{(2a-1)k_z} 
\left [ (1-a) \left ( \frac{k_z}{k_x} + \frac{5}{6} \frac{k_x}{k_z} \right ) + \frac{a}{2}\frac{k_x}{k_z} \right ]
\right \}, \hspace{3cm}
\\
-\frac{k_x}{k_z} \left [ 1 - f \frac{8\tilde\kappa^2(1-a) }{(2a-1)k_x k_z} \left ( \frac{k_z}{k_x} + \frac{5}{6}\frac{k_x}{k_z} \right )
\right ], \hspace{4cm} \\
\frac{4{\rm i} \tau \tilde\kappa^2 a k_x}{(2a-1)k_z^2} 
\left \{ 1 + \frac{4f (1-a) k_x}{(2a-1)^2 k_z} \left [ (1-a) \left ( \frac{k_z}{k_x} + \frac{5}{6}\frac{k_x}{k_z} \right ) + \frac{a}{2} \frac{k_x}{k_z} \right ]  \right . 
\left . + \frac{2f a}{2a-1} - \frac{2}{2a-1} \frac{\Delta k_z}{k_z} + \frac{\Delta k_x}{ak_x}
\right \}. 
\end{aligned}
\right \}
\end{equation}

iii) IW$^+$

\begin{equation}
\left \{
\begin{aligned}
\label{app_IW1_iwp}
1,\hspace{8cm} \\
\\
-\frac{{\rm i \tilde\kappa^2}}{k_z} \left \{ 1 - \frac{2f k_x}{k_z} 
\left [ \frac{(1-a)}{2a-1} \left ( \frac{k_z}{k_x} + \frac{5}{6} \frac{k_x}{k_z} \right ) + \frac{a}{2(2a+1)}\frac{k_x}{k_z} \right ] \right \}, \hspace{3cm} \\
\\
-\frac{k_x}{k_z} \left [ 1 - f \frac{8 \tilde\kappa^2 (1-a) }{(2a-1)k_x k_z} \left ( \frac{k_z}{k_x} + \frac{5}{6}\frac{k_x}{k_z} \right )  
\right ], \hspace{5cm} \\
-\frac{4{\rm i} \tau \tilde\kappa^2 a k_x}{(2a+1)k_z^2} 
\left \{ 1 - \frac{4f (a+1)}{(2a+1) } \frac{k_x}{k_z} \left [ \frac{(1-a)}{2a-1} \left ( \frac{k_z}{k_x} + \frac{5}{6}\frac{k_x}{k_z} \right ) + \frac{a}{2(2a+1)} \frac{k_x}{k_z} \right ]  \right . 
\left . + \frac{2f a}{2a+1} - \frac{2}{2a+1} \frac{\Delta k_z}{k_z} + \frac{\Delta k_x}{ak_x}
\right \}. 
\end{aligned}
\right \}
\end{equation}

At the wavenumber of IW$^{\prime\prime}_r$ the eigen-vectors of SDW, IW$^-$ and IW$^+$ are obtained, respectively,
from equations (\ref{app_IW1_sdw}), (\ref{app_IW1_iwm}) and (\ref{app_IW1_iwp})
by the replacements $a \to 1-a$ and $\Delta k_{x,z} \to -\Delta k_{x,z}$.

\section{Necessary dual eigen-vectors}
\label{dual_basis}

I) At the wavenumber of SDW$_r$.

i) SDW

\begin{equation}
\label{app_dual_SDW_sdw}
\left \{
\begin{aligned}
\frac{{\rm i} \tau k_x}{3} + \frac{8}{27} f \tau\, \frac{{\rm i} k_x (k_z^2 +8)}{k_z^2}, \\ 
-\frac{8\tau}{3} \frac{k_x}{k_z} - \frac{16}{27} f \tau\, \frac{ k_x (k_z^2 +26)}{k_z^3}, \\
- \frac{{\rm i} \tau k_x^2}{3 k_z} - \frac{8}{27} f \tau\, \frac{{\rm i} k_x^2 (k_z^2 +8)}{k_z^3}, \\
1 + \frac{8}{9} f\, \frac{k_x^2}{k_z^2} . \quad \quad \quad \quad
\end{aligned}
\right \}
\end{equation}

II) At the wavenumber of IW$^\prime_r$.

i) IW$^-$

\begin{equation}
\label{app_dual_IW1_iwm}
\left \{
\begin{aligned}
\frac{k_z^2}{8\tilde\kappa^2} + \frac{f}{\tilde\kappa^2} \frac{3a (44 + k_z^2) - (4a^2 +2) (20+k_z^2)}{24 (2a-1)^2}, 
\hspace{1.5cm}\ \\
\frac{{\rm i} k_z}{2\tilde\kappa^2} + \frac{{\rm i} f}{\tilde\kappa^2} \frac{(3a-1)(20+k_z^2) - 8 a^2 (2 + k_z^2)}{6 (2a-1)^2 k_z}, \hspace{1.5cm} \\ 
- \frac{k_x k_z}{8\tilde\kappa^2} - 
\frac{f}{\tilde\kappa^2} \frac{(4a^2+2)(k_x^4 - 26 k_x^2\tilde\kappa^2 + 48\tilde\kappa^4) - 3 a (k_x^4 - 52k_x^2\tilde\kappa^2 + 96\tilde\kappa^4)}
{24 k_x k_z (2a-1)^2}, \\
-{\rm i} \frac{f}{\tau} \frac{k_x}{4(2a-1)\tilde\kappa^2}.\hspace{3cm} 
\end{aligned}
\right \}
\end{equation}

At the wavenumber of IW$^{\prime\prime}_r$ the dual eigen-vector of IW$^-$ 
is obtained from equation (\ref{app_dual_IW1_iwm}) by the replacement $a \to 1-a$.

\section{Derivation of coupling coefficients}
\label{coupl_coeffs}

Let the eigen-vectors of the resonant triad, $\phi_{({\bf k})[1]}^i$, $\phi_{({\bf k}^\prime) [2]}^i$ and $\phi_{({\bf k}^{\prime\prime}) [2]}^i$,
consist of the following components
\begin{eqnarray}
\label{comp_eig}
&\phi_{({\bf k})i[1]} = \{ \hat u_x, \hat u_y, \hat u_z, \hat \delta \}^T, \nonumber \\
&\phi_{({\bf k}^\prime)i[2]} = \{ \hat u_x^\prime, \hat u_y^\prime, \hat u_z^\prime, \hat \delta^\prime \}^T, \nonumber \\
&\phi_{({\bf k}^{\prime\prime})i[2]} = \{ \hat u_x^{\prime\prime}, \hat u_y^{\prime\prime}, \hat u_z^{\prime\prime}, \hat \delta^{\prime\prime} \}^T,
\nonumber \\
\end{eqnarray}
explicitly given by the respective equations (\ref{app_SDW_sdw}), (\ref{app_IW1_iwm}) and 
(\ref{app_IW1_iwm}) with the replacements $a\to 1-a$ and $\Delta k_{x,z} \to - \Delta k_{x,z}$.

Similarly, let the corresponding dual eigen-vectors, 
$\tilde \phi_{({\bf k})i[1]}$, $\tilde \phi_{({\bf k}^\prime)i[2]}$ and $\tilde \phi_{({\bf k}^{\prime\prime})i[2]}$,
consist of the following components
\begin{eqnarray}
\label{comp_dual_eig}
&\tilde\phi_{({\bf k})i[1]} = \{ \tilde u_x, \tilde u_y, \tilde u_z, \tilde \delta \}^T, \nonumber \\
&\tilde\phi_{({\bf k}^\prime)i[2]} = \{ \tilde u_x^\prime, \tilde u_y^\prime, \tilde u_z^\prime, \tilde \delta^\prime \}^T, \nonumber \\
&\tilde\phi_{({\bf k}^{\prime\prime})i[2]} = \{ \tilde u_x^{\prime\prime}, \tilde u_y^{\prime\prime}, \tilde u_z^{\prime\prime}, 
\tilde \delta^{\prime\prime} \}^T, \nonumber \\
\end{eqnarray}
explicitly given by the respective equations (\ref{app_dual_SDW_sdw}), (\ref{app_dual_IW1_iwm}) and 
(\ref{app_dual_IW1_iwm}) with the replacement $a\to 1-a$.

The notations in RHS of equations (\ref{comp_eig}) and (\ref{comp_dual_eig}) 
are used in this Section only, in order to outline the derivation of equations (\ref{Q_1}) and (\ref{Q_2}).

Inspection of equations (\ref{C_1}), (\ref{C_2}) and (\ref{C_3}) shows that, to leading order in $\tau$ and $f$, the relevant terms can 
be arranged into combinations containing either 
\begin{equation}
\label{div_u_1}
D^{\prime\prime} \equiv k_x^\prime \hat u_x^{\prime\prime} + k_z^\prime \hat u_z^{\prime\prime}  = 
k_x \hat u_x^{\prime\prime} + k_z \hat u_z^{\prime\prime} =
f \tilde\kappa^2 \frac{8a}{1-2a} \frac{20\tilde \kappa^2 + k_z^2}{6 k_x k_z^2}
\end{equation}
or
\begin{equation}
\label{div_u_2}
D^{\prime} \equiv k_x^{\prime\prime} \hat u_x^{\prime} + k_z^{\prime\prime} \hat u_z^{\prime} = k_x \hat u_x^{\prime} + k_z \hat u_z^{\prime} =
f \tilde\kappa^2 \frac{8(1-a)}{2a-1} \frac{20\tilde \kappa^2 + k_z^2}{6 k_x k_z^2}.
\end{equation}
As expected, equation (\ref{div_u_2}) can be obtained from equation (\ref{div_u_1}) by the replacement $a \to 1-a$.
Note that $D^\prime,D^{\prime\prime} \sim O(f)$. 

In this way,
\begin{equation}
\label{N_Q_1}
Q_1 = 
-{\rm i} (\hat u_x^\prime \tilde u_x^* + \hat u_y^\prime \tilde u_y^* + \hat u_z^\prime \tilde u_z^* + \hat \delta^\prime \tilde \delta^* ) 
D^{\prime\prime}
- {\rm i} (\hat u_x^{\prime\prime} \tilde u_x^* + \hat u_y^{\prime\prime} \tilde u_y^* + 
\hat u_z^{\prime\prime} \tilde u_z^* + \hat \delta^{\prime\prime} \tilde \delta^* ) 
D^{\prime},
\end{equation}
where combinations in front of $D^\prime$ and $D^{\prime\prime}$ are taken to the zeroth order in $f$. 
Equation (\ref{N_Q_1}) shortly leads to equation (\ref{Q_1}).

Further on,
\begin{equation}
\label{N_Q_23}
Q_2 = 
-{\rm i} (\hat u_x \tilde u_x^{\prime\,*} + \hat u_y \tilde u_y^{\prime\, *} + \hat u_z \tilde u_z^{\prime\, *} + 
\hat \delta \tilde \delta^{\prime\, *} ) 
D^{\prime\prime}
+ {\rm i} ( \hat u_x^{\prime\prime \,*} \tilde u_x^{\prime\,*} + \hat u_y^{\prime\prime \,*} \tilde u_y^{\prime\, *} + 
\hat u_z^{\prime\prime \,*} \tilde u_z^{\prime\, *} ) (k_x^{\prime\prime} \hat u_x + k_z^{\prime\prime} \hat u_z),
\end{equation}
where the combination in front of $D^{\prime\prime}$ is taken to the first order in $f$, whereas the part of 
this equation standing after $D^{\prime\prime}$ yields zero to order of $f^2$. Equation (\ref{N_Q_23}) shortly leads to equation (\ref{Q_2}).
Derivation of $Q_3$ is identical to $Q_2$ with the replacement of superscripts $\prime \leftrightarrow \prime \prime$ in equation (\ref{N_Q_23}).


\end{document}